\documentclass{aa}
\newif\ifpdf\ifx\pdfoutput\undefined\pdffalse\else\pdfoutput=1\pdftrue\fi
\ifpdf
  \usepackage[pdftex]{graphicx}
\else
  \usepackage{graphicx}
\fi
\usepackage{natbib}
\usepackage[dvips]{color}
\usepackage{amsmath,amssymb}
\usepackage{txfonts}

\bibpunct{(}{)}{;}{a}{}{,} 

\def\rhod{\ensuremath{\rho_{\mathrm{d}}}}
\def\fdrag{\ensuremath{f_{\mathrm{drag}}}}
\def\alfa{\ensuremath{\langle\hat{\alpha}\rangle}}
\def\ssD{\ensuremath{_{\mathrm{D}}}}

\def\kms{\ensuremath{\,\mbox{km}\,\mbox{s}^{-1}}}
\def\kns{\ensuremath{\,\mbox{km}/\mbox{s}}}
\def\CtoO{\ensuremath{\varepsilon_{\mathrm{C}}/\varepsilon_{\mathrm{O}}}}
\def\drhog{\ensuremath{\rho_{\mathrm{d}}/\rho}}
\def\taud{\ensuremath{\tau_{\mathrm{d}}}}
\def\uinf{\ensuremath{u_{\infty}}}
\def\fcond{\ensuremath{f_{\mathrm{cond}}}}
\def\vdri{\ensuremath{v_{\mathrm{D}}}}
\def\qdrift{\ensuremath{q_{\mathrm{drift}}}}
\def\teff{\ensuremath{T_{\mathrm{eff}}}}
\def\deltaup{\ensuremath{\Delta\,u_{\mathrm{p}}}}
\def\subfst{\ensuremath{_{1\mathrm{st}}}}
\def\subsnd{\ensuremath{_{2\mathrm{nd}}}}
\def\upa{\ensuremath{\vartriangle}}
\def\dpa{\ensuremath{\triangledown}}
\def\mdotu{\ensuremath{\mbox{M}_{\sun}\,\mbox{yr}^{-1}}}
\def\mmdotv{\ensuremath{[10^{-6}\,\mbox{M}_{\sun}/\mbox{yr}]}}

\def\sistd{\ensuremath{\sigma_{\mathrm{s}}}}
\def\mmdot{\ensuremath{\langle\dot{M}\rangle}}
\def\muinf{\ensuremath{\langle u_{\infty}\rangle}}
\def\mfcond{\ensuremath{\langle f_{\mathrm{cond}}\rangle}}
\def\mdrhog{\ensuremath{\langle\drhog\rangle}}
\def\mtaud{\ensuremath{\langle\tau_{\mathrm{d}}\rangle}}
\def\mvdri{\ensuremath{\langle v_{\mathrm{D}}\rangle}}
\def\mdrad{\ensuremath{\langle r_{\mathrm{d}}\rangle}}

\definecolor{mygray}{gray}{0.4}
\def\boxfd{\ensuremath{\textcolor{mygray}{\blacksquare}}}
\def\triup{\ensuremath{\textcolor{mygray}{\blacktriangle}}}
\def\tridn{\ensuremath{\blacktriangledown}}
\def\bstar{\ensuremath{\star}}

\newcommand{\rSaHo}{Paper~I}
\newcommand{\rHoDo}{HD97}
\newcommand{\rHoJoLoAr}{HJLA98}
\newcommand{\rWiLBe}{W00}
\newcommand{\rKrGaSe}{KGS94}

\begin{document}
\title{Three-component modeling of C-rich AGB star winds}
\subtitle{II. The effects of drift in long-period variables}
\author{C.~Sandin\and{S.~H\"ofner}}
\institute{Department of Astronomy and Space Physics,
Uppsala University, Box 515, S-751 20 Uppsala, Sweden}
\offprints{C.~Sandin,\\\email{Christer.Sandin@astro.uu.se}}
\date{Received 2002/Accepted}

\abstract{We present three-component wind models for carbon rich pulsating AGB stars. In particular we study the effects of drift in models of long-period variables, meaning that the dust is allowed to move relative to the gas (drift models). In addition we investigate the importance of the degree of variability of the wind structures. The wind model contains separate conservation laws for each of the three components of gas, dust and the radiation field. We use two different representations for the gas opacity, resulting in models with different gas densities in the wind. The effects which we investigate here are important for the understanding of the wind mechanism and mass loss of AGB stars. This study is hereby a necessary step towards more reliable interpretations of observations. We find that the effects of drift generally are significant. They cannot be predicted from models calculated without drift. Moreover, the non-drift models showing the lowest mass loss rates, outflow velocities, and the smallest variability in the degree of condensation do not form drift model winds. The wind formation in drift models is, except for a few cases, generally less efficient and the mass loss consequently lower than in the corresponding non-drift models. The effects of drift are generally larger in the more realistic models using that representation of the gas opacity which results in lower densities. The outflow properties of these models are also -- for all cases we have studied -- sensitive to the period of the stellar pulsations. A check of the mass loss rates against a (recent) fit formula shows systematically lower values, in particular in the more realistic models with a low density. The fit is in its current form inapplicable to the new models presented here.
}

\authorrunning{C.~Sandin \& {S.~H\"ofner}}
\titlerunning{Three-component modeling of AGB star winds. II}
\maketitle

\section{Introduction}\label{sec:introduction}
Stellar winds of asymptotic giant branch (AGB) stars are believed to form as a consequence of an efficient dust condensation in the extended atmosphere of these stars. Dust grains absorb radiative momentum from the highly luminous star and are thereby pushed away from it. While moving away from the star, they collide with gas particles and the momentum is transferred to the rest of the matter, forming the stellar wind. Mass loss and its history is not only important to the AGB star itself. It is also relevant for its neighborhood, the formation of the circumstellar envelope and in a longer perspective, the galactic chemical evolution \citep[e.g.\@][]{Wi:00}.

Since \citet{Bo:88} published his pioneering paper on time-dependent wind formation in cool M-type giants considerable progress has been achieved with models describing different physical aspects in more detail. \citet{FlGaSe:92} created a C-rich wind model using a self-consistent description of dust formation. \citet{FlGaSe:95} and \citet{HoFeDo:95} studied models without pulsations and described the dust-induced (external) kappa mechanism. The influence of molecular opacities for the wind driving efficiency were studied by \citet[henceforth {\rHoJoLoAr}]{HoJoLoAr:98} and \citet{HeWiSe:00}. The models in this article differ from the ones presented in the papers mentioned above as drift between gas and dust is taken into consideration. So far this phenomenon has almost exclusively been studied in stationary winds (e.g.\@ \citealt{KrGaSe:94} (henceforth {\rKrGaSe}), \citealt{KrSe:97}, or \citealt{LiLaBe:01}) or non-pulsating stars (e.g.\@ \citealt{SiIcDo:01} or \citealt{SaHo:03} (henceforth {\rSaHo})).

In the first article in this series, {\rSaHo}, we presented a method on how to describe drift in a C-rich wind model. The work is based on the model presented in \citet{HoFeDo:95} and both aspects of physical and numerical details were covered therein. The main objective in {\rSaHo} was to study the basic effects of drift on the wind structure; by comparing the results with position coupled (PC; i.e.\@ non-drift) models. In particular we studied different approximations of the term describing the collisional gas-dust momentum transfer, the drag force. Processes that could make the interpretation of the results more difficult were excluded. Consequently we did not describe stellar pulsations in the models of {\rSaHo}. And the positive effects of atmospheric levitation by pulsation were hereby excluded, requiring a high carbon/oxygen ratio and a luminous central star.

The main result of {\rSaHo} was that the dust in drift models tends to accumulate to the dense regions behind shocks, which form as a consequence of the dust-induced kappa-mechanism. This accumulation of dust does not occur in PC models. While the differences were obvious in the spatial data we could not find any \emph{certain} changes -- between drift and PC models -- in quantities which are temporally averaged at the outer boundary, such as the average mass loss rate and the average terminal velocity. The reason why we considered the averaged results of the drift models uncertain was that the numerical accuracy of the momentum flux terms in the drift models never was better than first order in {\rSaHo}.

The final goal is to replace the PC models with drift models, if it turns out to be necessary for a specific purpose. However, before we can decide on this matter it is necessary to find out what effect the drift has on the models, and whether there are clear correlations of the drift to other quantities. Is it possible to describe the effects on the wind structure without explicitly including the drift in the calculations? Such a treatment would be time-economically valuable since the larger variability in the dust component caused by drift inevitably slows down the calculations.

With the work presented in this article we improve the model presented in {\rSaHo} further. The numerical accuracy in the flux terms of the partial differential equations is now of second order for all models. We here include a description of stellar pulsations to study long-period variables. A commonly used representation of the gas opacity in AGB wind models is a constant number introduced by \citet{Bo:88}. For comparison with the corresponding models we calculate models using this representation too. However, since its use results in unrealistic density structures (cf.\@ {\rSaHo}, Sect.~4.1) we also calculate models based on molecular data (using Planck mean absorption coefficients; {\rHoJoLoAr}). The selection of sets of model parameters is based upon a sample where we expected that drift effects could be important; the models in the sample are most closely related to the models presented by \citet[henceforth {\rHoDo}]{HoDo:97} and {\rHoJoLoAr}. The only related study of time-dependent dust-driven winds -- we are aware of -- that self-consistently treats dust formation and drift is presented by \citet[ch.~4]{YSi:01}. However, the models presented therein are calculated using assumptions that are different from ours in several respects (cf.\@ {\rSaHo}, Sect.~4.2).

The physical assumptions and the model characteristics of the current work, compared to {\rSaHo}, are specified in Sect.~\ref{sec:physassum}. The parameter-relations and the criteria we use in the selection of the sets of model parameters are given in Sect.~\ref{sec:detmodel}, which also covers the modeling method. The results of the study are presented and discussed in two separate sections. The output of the models in terms of average properties of the wind and their inter-relations, are given in Sect.~\ref{sec:results}. Four topics associated to the new models and the effects of drift are discussed in further detail in Sect.~\ref{sec:discuss}. The topics are: the influence of the piston period on the properties of the wind; the importance of a ``temporal variability'' for the formation of winds and for discerning winds of different character; a discussion on the comparison with a mass loss fit formula; and the influence of heating by drift on the wind structure. The conclusions are given in Sect.~\ref{sec:conclus}.

\section{Physical assumptions and model characteristics}\label{sec:physassum}
The stellar wind can physically be separated into three different interacting components. All but a small fraction of the matter is assumed to form the gas component. The remaining matter, present in dust grains, forms the dust component. The third component is the radiation field.

\begin{table}
\caption{Glossary of used symbols and units}
\label{app:symbols}
\begin{tabular}{lll}\hline\hline\\[-1.8ex]
sym. & unit & description\\[1.0ex]\hline\\[-1.8ex]
  $\rho$           & $\mbox{g}\,\mbox{cm}^{-3}$      & gas density \\
  {\rhod}          & $\mbox{g}\,\mbox{cm}^{-3}$      & dust density \\
  $n_{\mathrm{g}}$ & $\mbox{cm}^{-3}$                & gas number density \\
  $n_{\mathrm{d}}$ & $\mbox{cm}^{-3}$                & dust number density \\
  $u$              & $\mbox{cm}\,\mbox{s}^{-1}$      & gas velocity \\ 
  $v$              & $\mbox{cm}\,\mbox{s}^{-1}$      & dust velocity \\
  $J$              & $\mbox{erg}\,\mbox{cm}^{-2}\,\mbox{s}^{-1}$  & zeroth moment of the \\
                                                   & & radiation field \\
  $K_j$            & cm$^{-3}$                       & moments of the grain size \\
                                                   & & distribution; $0\le j\le3$ \\
  {\vdri}          & $\mbox{cm}\,\mbox{s}^{-1}$      & drift velocity \\ 
  $S\ssD$          &                                 & speed ratio \\
  $C\ssD$          &                                 & drag coefficient \\
  $\varepsilon$    &                                 & fraction of specular collisions \\
  $\sigma$         & cm$^{2}$                        & gas-dust geom. cross section \\
  {\fdrag}         & $\mbox{g}\,\mbox{cm}^{-2}\mbox{s}^{-2}$ & drag force \\
  {\qdrift}        & $\mbox{erg}\,\mbox{cm}^{-3}\mbox{s}^{-1}$ & heating by drift \\
  {\mdrad}         & cm                              & mean grain radius\\[1.0ex]
  $P$              & $\mbox{dyn}\,\mbox{cm}^{-2}$    & gas pressure \\
  $T_{\mathrm{g}}$ & K                               & gas temperature \\
  $T_{\mathrm{d}}$ & K                               & dust temperature \\
  $S_{\mathrm{g}}$ & $\mbox{erg}\,\mbox{cm}^{-2}\,\mbox{s}^{-1}\,\mbox{ster}^{-1}$ & source function of the gas\\
  $\kappa_{\mathrm{g}}$ & $\mbox{g}^{-1}\,\mbox{cm}^{2}$ & (gray) gas opacity \\
  $\kappa_{\mathrm{d}}$ & $\mbox{g}^{-1}\,\mbox{cm}^{2}$ & (gray) dust opacity\\[0.5ex]
  $c$              & $\mbox{cm}\,\mbox{s}^{-1}$      & speed of light \\
  $k_{\mathrm{B}}$ & $\mbox{erg}\,\mbox{K}^{-1}$     & Boltzmann constant \\
  $G$              & $\mbox{dyn}\,\mbox{g}^{-2}\,\mbox{cm}^{2}$ & gravitational constant \\
  $\sigma_{\mathrm{B}}$ & $\mbox{erg}\,\mbox{cm}^{-2}\,\mbox{s}^{-1}\mbox{K}^{-4}$ & Stefan-Boltzmann constant \\
                   &                                 & \\
  \multicolumn{3}{l}{Quantities specified at the outer boundary}\\[1.0ex]\hline\\[-1.8ex]
  $\dot{M}$        & {\mdotu}                        & mass loss rate \\
  {\uinf}          & $\mbox{cm}\,\mbox{s}^{-1}$      & terminal velocity\\
  {\fcond}         &                                 & degree of condensation\\
  {\taud}          &                                 & dust optical depth\\
  {\sistd}         &                                 & standard deviation\\
                   &                                 & \\
  \multicolumn{3}{l}{Grain properties}\\[1.0ex]\hline\\[-1.8ex]
  \multicolumn{3}{l}{$K_0=n_{\mathrm{d}}$}\\
  \multicolumn{3}{l}{$K_1/K_0=\mdrad/r_0$;\;$r_0\,[\mbox{cm}]$ --- monomer radius}\\
  \multicolumn{3}{l}{$K_2/K_0=\langle A\rangle/4\pi r_0^2$;\;$\langle A\rangle\,[\mbox{cm}^2]$ --- mean grain surface area}\\
  \multicolumn{3}{l}{$K_3/K_0=\langle N\rangle$;\;$\langle N\rangle$ --- mean grain size}\\[1.0ex]\hline\\[-1.8ex]
\end{tabular}
\end{table}


In our wind model each of the three components is described by coupled conservation equations covering radiation hydrodynamics, time-dependent dust formation and interaction between all components; we call it the RHD3-system (Radiation HydroDynamics, Dust and Drift). The physical system and how it is solved is described thoroughly in {\rSaHo}. In short the equation of continuity, the equation of (internal) energy and the equation of motion describe the gas component. The dust component is assumed to be made of spherical particles in the form of amorphous carbon and is described with the dust equation of motion and four moment equations treating the formation and destruction of dust grains; the moments ($K_0$-$K_3$) describe certain properties of the grain size distribution function. Currently dust formation is treated considering the processes of nucleation, growth, evaporation and chemical sputtering. The radiation field is represented by two moment equations of the radiation transfer equation. The two moments are the frequency-integrated zeroth and first moments of the radiation intensity, representing the radiative energy density and the radiative energy flux, respectively. All symbols and quantities used in this article are (for a quick reference) given in Table~\ref{app:symbols}.

In the three following subsections we look at processes where the physics used in {\rSaHo} is improved further.

\subsection{Treatment of the gas-dust interaction}\label{sec:physgdin}
In this article we primarily present and discuss the results of drift models -- i.e.\@ models where the dust is allowed to drift with respect to the gas -- but for comparison we also present PC models that \emph{a priori} assume complete momentum coupling (and a tight mechanical coupling of the dust to the gas, preventing drift). In drift models the drift velocity $v_{\mathrm{D}}$ is defined by,
\begin{eqnarray}
\vdri=v-u
\end{eqnarray}
where $v$ is the dust velocity and $u$ the gas velocity. All three processes of momentum, energy and mass transfer between the gas and dust phases are affected by drift.

The description of \emph{the momentum transfer between the gas and the dust} is contained in the drag force which is calculated taking care of both the geometrical and thermal properties of the dust \emph{and} the thermal properties of the gas,
\begin{eqnarray}
f_{\mathrm{drag}}=\sigma(K_0,K_1)\rho n_{\mathrm{d}}\frac{v_{\mathrm{D}}^2C_{\mathrm{D}}(v_{\mathrm{D}},T_{\mathrm{g}},T_{\mathrm{d}},\varepsilon)}{2}\,.
\label{eq:dragforce}
\end{eqnarray}
Here $\sigma$ is the gas-dust geometrical cross section; $K_0$ and $K_1$ are the first two moments of the grain size distribution function; $\rho$ is the gas density; $n_{\mathrm{d}}$ is the dust number density (i.e.\@ $K_0$); $C\ssD$ the drag coefficient; $T_{\mathrm{g}}$ the gas kinetic temperature; $T_{\mathrm{d}}$ the dust temperature; and $\varepsilon$ the fraction of specular gas-dust collisions ($0\!\le\!\varepsilon\!\le\!1$). The drag coefficient we adopt is $C_{\mathrm{D}}^{\mathrm{LA}}$ given in Eq.~(23) in {\rSaHo}. The exact form of the gas-dust geometrical cross section, the dust properties, and the assumptions used for the dust temperature are given in {\rSaHo}. The term in $C\ssD$ that accounts for diffusive collisions becomes significant at low drift velocities and large dust/gas temperature ratios, see Eq.~(10) in {\rSaHo}. However, in Fig.~4c (in Sect.~4.3 {\rSaHo}) it is seen that the same term never accounts for more than 40\% of $C\ssD$ and the calculated wind structure is more or less independent of the type of collisions. Consequently, with physical conditions similar to those described in {\rSaHo} we assume fully specular collisions ($\varepsilon=1$).

\emph{The transfer of internal energy} (dependent on drift) between gas and dust can be split in two terms. On the one hand heat is transferred to gas particles that are accommodated on the dust grain surface in diffusive collisions. For a complete derivation of this term see e.g.\@~\citet[and the equations leading to Eq.~(7)]{Sc:63}; {\rKrGaSe} (Eq.~(14)) discuss an approximative form. On the other hand energy of the bulk motion is converted into internal (thermal) energy as gas particles which preferentially come from one direction (as in the case of drift) are reflected in random directions when hitting a dust particle. Calculations involving the details of a two-body collision show that this heating (by drift) is given by, 
\begin{eqnarray}
\qdrift=\frac{1+\varepsilon}{2}\vdri\fdrag\label{eq:qdrift}
\end{eqnarray}
(e.g.\@ {\rKrGaSe} and \citealt{Dr:80,Dr:86}). {\rKrGaSe} found heating by drift to be a significant factor in the energy balance of the gas in stationary drift models with a simplified description of the radiation field and the dust. In our time-dependent models we find that the impact of this heating on the wind structure is low; cf.\@ Sect.~\ref{sec:discussh}. In contrast to {\rSaHo}, where {\qdrift} was not included in any calculations, it is in this article included in all calculations presented in Sects.~\ref{sec:results} \& \ref{sec:discussp}.

\emph{The mass transfer} (represented by dust formation and destruction) is also affected by drift, partly through modified grain growth efficiencies, partly in non-thermal sputtering that occurs frequently if the drift velocity exceeds about 40\kms~\citep[cf.\@][and references therein]{KrSe:97}. In the results of {\rSaHo} we found that a significant fraction of the dust in drift models accumulates to the regions behind gas shocks, the drift velocity was also found to stay below $30\kms$. The latest calculations presented in this article do not significantly alter the validity of the conditions behind these assumptions. Currently we neither include sputtering due to drift nor modifications of the dust formation rates to account for drift. The plan is, however, to include them in the calculations and study their influence on the wind structure closely in the near future.

\subsection{Treatment of the gas-radiation field interaction}\label{sec:gasradin}
Starting with~\citet{Bo:88} it has been customary to assign the gas opacity a (low) constant number as a primary estimate,
\begin{eqnarray}
\kappa_{\mathrm{g}}=2\times10^{-4}\,[\mbox{cm}^2\,\mbox{g}^{-1}]\,.\label{eq:physkapc}
\end{eqnarray}
It was done on the basis that reliable opacities at that time were not available and the value given here was thought to be representative in the region of the wind where the gas opacity was believed to play a r{\^o}le. To mention just a few, \citet{Bo:88}, \citet{FlGaSe:92}, \citet{FlGaSe:95}, \citet{HoFeDo:95}, {\rHoDo}, \citet{WiFlLBeSe:97}, \citet[henceforth {\rWiLBe}]{WiLBe.:00}, and {\rSaHo} all present time-dependent dust-driven wind models that adopt this opacity.

The constant value on the gas absorption coefficient ($\kappa_{\mathrm{g}}$) is independent of both the thermodynamical conditions and the chemical composition and its use results in unrealistically large densities in the photospheric layer. It was replaced with Planck mean absorption coefficients based on the SCAN molecular data \citep{Jo:97} by {\rHoJoLoAr} and \citet{HeWiSe:00}. This improved description of the gas opacity avoids the extra ``free parameter'' (given in Eq.~(\ref{eq:physkapc})), is a function of the chemical composition, and allows for the calculation of a more realistic density structure.

The difference between structures of models using either of the two described opacities is in general significant, and the resulting winds show very different behaviors. In this article we both calculate models using a constant gas opacity (constant-opacity models) -- to compare with the results of previous models in the literature -- and models using an opacity based on Planck mean absorption coefficients (Planck mean models). 

\subsection{Modeling stellar pulsations}\label{sec:physpuls}
To describe the effects of stellar pulsations on the atmosphere we use a sinusoidal radially varying inner boundary, located at about $0.91\,R_*$ (above the region where the $\kappa$-mechanism supposedly originates). Since the radiative flux is assumed to be constant through the innermost mass shell the radial variation implies a varying luminosity as well. The above description of the pulsations is not self-consistent in the sense of a pulsation model, but it provides the wind with the dynamical effects of pulsations, i.e.\@, a levitated atmosphere and strong shocks (cf.\@ {\rHoJoLoAr} and references therein). The period of the piston is in the previous work of e.g.\@ {\rHoDo} and {\rHoJoLoAr} specified through a period-luminosity (P-L) relation derived and fitted for M-type Miras of periods $P$ shorter than 420\,days by~\citet{FeGlWhCa:89},
\begin{eqnarray}
\log P=2.5/3\cdot\log\left(L/L_{\sun}\right)-1.84/3
\label{eq:physpulsa}
\end{eqnarray}
where $P$ is given in days. The piston amplitude is a ``free parameter'', see Sect~\ref{sec:detmodel}.

The influence of different pulsation periods on the wind structure is studied by e.g.\@ {\rWiLBe} and {\rHoDo}. The former authors concluded that a long period allows for a stronger levitation of the atmosphere and more favorable conditions for dust formation. However, these effects are counteracted by a more efficient radiation pressure on the dust that dominates the dynamics and cancels the improved efficiency. Similarly {\rHoDo} found the effects of altered periods to be insignificant in two of their PC models.

We want to see if this conclusion is valid for drift models and therefore complement the models calculated using Eq.~(\ref{eq:physpulsa}) with a more recent P-L relation fitted by \citet[Eq.~(2)]{GrWh:96},
\begin{eqnarray}
\log P=2.5/2.59\cdot\log\left(L/L_{\sun}\right)-2.71/2.59\,.
\label{eq:physpulsc}
\end{eqnarray}
Equation~(\ref{eq:physpulsc}) is derived for carbon Miras with periods in the range 150-520\,days. Hence there is still a need to extrapolate to longer periods. \citet{WiFlLBeSe:97} point out that these extrapolated periods are possibly overestimated. The tendency towards overestimated periods can also be seen in \citet[Fig.~3]{GrvLoWh.:00} where (O-rich and) C-rich stars with longer periods are measured. These authors point out that both P-L relations in our discussion are limited to periods of less than 400-500 days because they are based on a sample of optically visible stars with shorter periods. The extrapolated periods ($>520$ days) given by the P-L relation of \citet{GrWh:96} should thus be considered an upper limit.

\section{Details of the modeling}\label{sec:detmodel}
The equations of the RHD3-system are discretized in the volume-integrated conservation form on a staggered mesh. The spatial discretization of the advection terms is always of second order~\citep{vLe:77} in this article, contrasting {\rSaHo} where it was never higher than first order in the drift models. The RHD3-system of equations is solved implicitly using a Newton-Raphson algorithm where the Jacobian is inverted by the Henyey method (cf.\@ {\rSaHo} for details).

\begin{table}
\caption{Model parameters ($L_*$, {\teff} and $P$) of both constant-opacity models (prefix ``R'') and Planck mean models (prefix ``P''). Columns 5 \& 6 give the stellar radius, calculated from $L_*$ and {\teff} (Eq.~(\ref{eq:teff})), and the gravitational acceleration at the location of the photosphere. The last column shows the fraction of the stellar mass contained in the model domain of the initial model for constant-opacity models and Planck mean models (the latter averaged over the carbon/oxygen ratios used), respectively. The stellar mass $M_*$ is set to $1.0\,M_{\sun}$ in all models. The remaining two (out of totally six) model parameters, the carbon/oxygen ratio {\CtoO} and the piston velocity amplitude {\deltaup} are specified in Tables~\ref{tab:resultcg}-\ref{tab:discusspr} (Sects.~\ref{sec:results} \& \ref{sec:discussp}).}\label{tab:detmodel}
\begin{tabular}{lccccccc}\hline\hline\\[-1.8ex]
\multicolumn{1}{c}{model} & $L_*$        & {\teff} & $P$   & \!\! &
 $R_*$        & $\log g_*$ & $M_{\mathrm{e}}/M_*$ \\
\multicolumn{1}{c}{R-/P-} & $[L_{\sun}]$ & $[\mbox{K}]$       & $[\mbox{d}]$ & \!\! &
 $[R_{\sun}]$ &            & $[\%]$ \\[1.0ex]\hline\\[-1.8ex]
07F & $7.0\times10^3$ & 2880 & 390 &\!\!& 336 & -0.61 & 13; 0.16 \\[0.5ex]
10F & $1.0\times10^4$ & 2790 & 525 &\!\!& 428 & -0.82 & 17; 0.16 \\
10G & $1.0\times10^4$ & 2790 & 653 &\!\!& 428 & -0.82 & 17; 0.16 \\[0.5ex]
13F & $1.3\times10^4$ & 2700 & 650 &\!\!& 521 & -0.99 & 20; 0.19 \\
13G & $1.3\times10^4$ & 2700 & 841 &\!\!& 521 & -0.99 & 20; 0.19 \\[1.0ex]\hline
\end{tabular}
\end{table}


\subsection{Selecting model parameters}\label{sec:modelp}
The model is determined by four stellar parameters and two parameters defining the piston, there are thus in total six parameters. The parameters are, the stellar mass $M_*$, the stellar luminosity $L_*$, and the effective temperature {\teff} defining the photosphere of the initial hydrostatic model. All abundances but that of the carbon are assumed solar; the carbon abundance is specified through the carbon/oxygen ratio \CtoO. Furthermore the period $P$ and the piston velocity amplitude {\deltaup} define the piston.

The purpose of this article is to study certain issues connected to drift that we believe are important for realistic wind structures. Therefore we intentionally select combinations of model parameters using established parameter relations. The sets of model parameters in this study, given in Table~\ref{tab:detmodel}, are mainly selected from the models given in {\rHoDo} and {\rHoJoLoAr}. Note that constant-opacity models are given the prefix ``R'' in the model name, and Planck mean models the prefix ``P''. In selecting model parameters we intentionally leave out the least luminous models with the less massive outflows. These models are \emph{a priori} not expected to form winds when adopting drift. We restrict the parameter space by not varying the stellar mass $M_*$ which is set to $1.0\,M_{\sun}$ in all models. The piston amplitude {\deltaup} and the carbon/oxygen ratio {\CtoO} are currently considered ``free parameters''; the values we use are specified in the model names given in the tables in the results section (Sect.~\ref{sec:results}).

In the models presented here it is the dust that initiates and drives the stellar wind. The dust formation zone, and hereby the wind formation zone, is always inside the model domain. An inflow of matter through the inner boundary which is located well below the photosphere (typically at about $0.9\,R_*$) is not allowed. One consequence of the lower density present in the Planck mean models, compared to constant-opacity models, is that a smaller fraction of the stellar mass is located in the atmosphere (compare the values in the last column in Table~\ref{tab:detmodel}). Consequently it is currently not possible to model a long-term wind evolution since the model domain is depleted quickly of material.

The effective temperature is specified using the radius-mass-luminosity relation -- derived from an evolutionary model -- given by \citet{Ib:84}\footnote{Equation~\ref{eq:ibenlmr} is based on old opacities resulting in overestimated effective temperatures, the newer relations presented by e.g.\@ \citet{WaGr:98} may be a better choice since they are based on more recent opacities.},
\begin{eqnarray}
R=312\,\left(\frac{L}{10^4}\right)^{0.68}\left(\frac{M}{1.175}\right)^{-0.31S}\left(\frac{Z}{0.001}\right)^{0.088}\left(\frac{l}{H_{\mathrm{p}}}\right)^{-0.52}\label{eq:ibenlmr}
\end{eqnarray}
where the stellar radius $R$, the stellar luminosity $L$ and the stellar mass $M$ are given in solar units. The remaining values in Eq.~(\ref{eq:ibenlmr}), the metalicity $Z=0.02$ and the ratio of the mixing length to the pressure scale height $(l/H_{\mathrm{p}})=0.90$ are taken from \citet{BoWi:91}. $S$ is identically 0 unless $M\ge1.175$ when $S=1$. The effective temperature is given by,
\begin{eqnarray}
\sigma_{\mathrm{B}}\teff=L_*/\left(4\pi R_*^2\right)\label{eq:teff}
\end{eqnarray}
where $\sigma_{\mathrm{B}}$ is the Stefan-Boltzmann constant.

The period $P$ is chosen according to the P-L relations mentioned in Sect.~\ref{sec:physpuls}. Model names with an ``F'' adopt the P-L relation derived by \citet{FeGlWhCa:89} (F-models), and those with a ``G'' that derived by \citet{GrWh:96} (G-models). The relation by \citeauthor{FeGlWhCa:89} is used in all models presented in Sect.~\ref{sec:results}. Results using different periods, represented by the G-models, are discussed in Sect.~\ref{sec:discussp}.

\subsection{Modeling procedure}\label{sec:modelpro}
The modeling procedure is the same for all models we present and is as follows. Each wind model is started from a hydrostatic dust-free initial model where the outer boundary is located at about $2\,R_*$. All five dust equations, i.e.\@ the dust equation of motion and the four dust moment equations, are switched on at the same time. Dust starts to form whereby an outward motion of the dust and the gas is initiated. The expansion is followed by the grid to about $25\,R_*$, where the outer boundary is fixed allowing outflow. The model evolves for about $50$-$200\,P$ in the case of drift models, and for about $400$-$900\,P$ in the case of PC models. The shorter evolutionary times for the drift models are due to the generally larger variations of the dust quantities (in particular) and the resulting smaller time steps in these models (also see Sect.~4.2 in {\rSaHo}). One may argue that the terminal velocities determined at our outer boundary which is located relatively close to the star may be underestimated. However, our experience shows that this does not appear to be the case\footnote{Tests have been performed (with previous versions of the code), where the outer boundary of the model is located at larger distances from the star. These tests have shown that the uncertainties in the terminal velocity caused by the location of the outer boundary are small compared to the uncertainties in the calculation of the time-averaged values.}.

The massive envelopes of constant-opacity models prevent depletion in these models. The density is on the contrary lower, and the envelope thereby less massive, in the Planck mean models. As a consequence the calculated quantities of the latter models are averaged over a time-period shorter than one where the depletion becomes relevant (cf.\@ Sect.~\ref{sec:resultpg}).

\subsection{Comparison of numerically improved non-pulsating wind models with old models}\label{sec:nonpulsa}
In {\rSaHo} we calculated drift models without stellar pulsations using a $1^{\mathrm{st}}$ order spatial discretization of the advection terms. In this subsection we compare those models (Table~2 in {\rSaHo}; drift{\subfst} models) with improved drift models that use a $2^{\mathrm{nd}}$ order advection (drift{\subsnd} models), Table~\ref{tab:modcomp}. Like in {\rSaHo} none of the improved models simulate stellar pulsations and for comparison the heating term {\qdrift} is not included. Considering the type of winds formed in the drift{\subfst} models, it was found that they were all irregular, but so were the PC models that used first order advection. The new drift{\subsnd} models are different in that they show much less irregularity and have shorter time-scales in the variational patterns. They are, however, all still irregular wind models.

\begin{table}
\caption{Model comparison of drift{\subfst} models (Cols.~2,3) vs.\@ drift{\subsnd} models (Cols.~4,5); completing Table~2 in {\rSaHo} (where the properties for the drift{\subfst} models are taken). Only those models that give a wind are shown. All wind models show an irregular variability. Models that are not run for a longer time are marked with parentheses. See Sect.~\ref{sec:nonpulsa} for further information.}
\label{tab:modcomp}
\begin{tabular}{cccccc}\hline\hline\\[-1.8ex]
      &\multicolumn{2}{c}{drift{\subfst} models}&&
        \multicolumn{2}{c}{drift{\subsnd} models}\\
model & $10^6\cdot$\mmdot &\muinf     &&$10^6\cdot$\mmdot&\muinf\\
       & $[\mdotu]$        &$[\!\kms]$ &&$[\mdotu]$       &$[\!\kms]$\\[1.0ex]\hline\\[-1.8ex]
A27 & (6.6) & (40) && (4.7) & (40) \\[0.5ex]
B21 & (11)  & (28) && (7.5) & (30) \\
B22 & (7.8) & (28) && (7.6) & (32) \\[0.5ex]
C18 & 15    & 21   && 12    &  23 \\
C19 & 14    & 24   && 11    &  26 \\
C20 & 13    & 25   && 11    &  28 \\[0.5ex]
D16 & 43    & 27   && 38    &  28 \\[1.0ex]\hline\\[-1.8ex]
\end{tabular}
\end{table}


While the measured temporally averaged terminal velocities are more or less the same when comparing drift{\subsnd} and drift{\subfst} models, the mass loss rates differ. The drift{\subsnd} model A27 has an average mass loss rate that is much lower than that of the drift{\subfst} model and is about the same as the value of the corresponding PC model. All other models have average mass loss rates in the range of $72$-$80\%$/$68$-$97\%$ of the corresponding PC/drift{\subfst} models (with a larger uncertainty for the shorter evolved B21 and B22 models).

The conclusion is that with the more accurate discretization of the drift{\subsnd} models, compared to the drift{\subfst} models, we find clear changes caused by drift in the wind structure of certain models. However, since the properties of both types of model winds are of the same order of magnitude and the dust still accumulates to the regions behind shocks, there is a rough qualitative agreement between drift{\subfst} and drift{\subsnd} models. Nevertheless a more accurate discretization is always preferred.

\section{Results}\label{sec:results}
In this section we present and compare the results of new drift- and PC models which include the effects of stellar pulsations. A discussion of the consequences of the results is given in Sect.~\ref{sec:discuss}.

All quantities given in both tables in this section (and in Table~\ref{tab:discusspr} in Sect.~\ref{sec:discussp}) are temporal means calculated at the (fixed) outer boundary. To make an estimate of the ``degree of variability'' for each model and quantity we calculate a standard deviation, see Sect.~\ref{sec:discussv} for a discussion of its relevance and implications. Note that while all models are calculated for a total time-interval which we believe is long enough to find reliable average quantities, i.e.\@ $50$-$400\,P$ depending on the model, we do not study long-term variations of the order of several hundred to thousands of years. Such a study requires total time-intervals on the order of thousands of periods and longer.

Depending on the choice of a time-interval used in the calculation of the average quantities slightly different values emerge. Since a majority of the models in this study are irregular it is difficult to determine general conditions for the exact choice of such a time-interval, instead we choose them as long as possible (see, however, Sect.~\ref{sec:resultpg}). We mention at this point that the time-interval used for the determination of the average drift velocity in our models necessarily is different, cf.\@ Sects.~\ref{sec:resultcg_drift}~\&~\ref{sec:resultpg_drift}.

The two different kinds of gas opacities we apply give intrinsically different wind structures and we therefore present the results of each of these two models in separate subsections. We first study the results of constant-opacity models and thereafter the results of (and a comparison with) the Planck mean models. To illustrate differences we have chosen to put results of different nature in the same figure. To aid the further discussion we here give the following symbol key for these figures (Figs.~\ref{fig:results_ratios}-\ref{fig:results_alfa}, \ref{fig:discuss_dev}, \ref{fig:discussm}): PC models are always represented by \emph{open symbols}, and drift models by \emph{filled symbols}. Moreover constant-opacity models are represented by circular symbols, stars, and downwards pointing triangles ($\circ,\oplus,\circledcirc,\bullet,\bstar,\tridn$), and Planck mean models by rectangular symbols and upwards pointing triangles ($\Box,\boxplus,\boxfd,\triup$). Planck mean drift models are plotted in \textcolor{mygray}{gray} to distinguish them more clearly from constant-opacity drift models. Triangular symbols represent models with a varied period $P$. Further details are given in the respective tables and figures.

\subsection{Results of constant-opacity models}\label{sec:resultcg}
The results of the constant-opacity models are given in Table~\ref{tab:resultcg}. PC models are given in the upper part and drift models in the lower part; note that we adopt the same model name for PC models and drift models with the same model parameters. The standard deviation {\sistd} of the average values is specified for the first five quantities (but not for the drift velocity, cf.\@ Sect.~\ref{sec:resultcg_drift}). We give the actual dust/gas density ratio in Col.~8 (instead of calculating it from the degree of condensation like in {\rHoDo} \& {\rHoJoLoAr}, which only works correctly for PC models).

\begin{table*}
\caption{Model quantities averaged at the outer boundary for constant-opacity models (cf\@. Sect.~\ref{sec:resultcg}). The models are named by adding a suffix to the respective model name in Table~\ref{tab:detmodel}. The suffix is a combination of two model parameters, the piston velocity amplitude {\deltaup} (U$n$, $n$ in {\kms}) and the carbon/oxygen ratio {\CtoO} (C$nn$, $nn$ is $10\cdot{\CtoO}$). The last but two column, $T_{\mathrm{tot}}$, gives the total time-interval of each model calculation. The numbers associated with models run for a shorter time-interval are less reliable than the others as the means are taken over short time intervals. The last but one column gives the type of wind: {\bf i}, irregular wind; $l${\bf p}, periodic wind; $l${\bf q}, quasi-periodic wind; {\bf t}, transition model; {\bf ---}, no wind. $l$ ($\in\mathbb{N}$) shows the (multi-)periodicity of dust shell formation in the unit of the piston period $P$. A tilde (e.g.\@ `2\~q') in the last but one column indicates a correspondence with the characterization only during a part of the calculated time-interval. The values shown in bold face for the PC models indicate that the value differs from the corresponding value given by {\rHoDo} (by $\ge\!10\%$). The symbols in the last column show whether the respective drift models show an increased/decreased (\upa/\dpa) mass loss rate, or new wind ($\circledast$) when compared to the corresponding PC model. For the PC models an `H' indicates that the same model parameters were used in a model in {\rHoDo}. The symbol printed in subscript in the last column indicates how the respective (differing) model is illustrated in Figs.~\ref{fig:results_mdot}, \ref{fig:results_alfa}, \ref{fig:discuss_dev} \& \ref{fig:discussm}.}
\label{tab:resultcg}
\begin{tabular}{lr@{\ \ }rcr@{\ \ }rcr@{\ \ }rcr@{\ \ }rcr@{\ \ }rrrrl}\hline\hline\\[-1.8ex]
   model & \multicolumn{2}{c}{\mmdot} &&
           \multicolumn{2}{c}{\muinf} &&
           \multicolumn{2}{c}{\mfcond} &&
           \multicolumn{2}{c}{\mdrhog} &&
           \multicolumn{2}{c}{\mtaud} &
           \multicolumn{1}{c}{\mvdri} &
           \multicolumn{1}{c}{$T_{\mathrm{tot}}$} &
           type \\
           \multicolumn{3}{r}{\mmdotv} &&
           \multicolumn{2}{c}{$[\!\kns]$} &&
           \multicolumn{2}{c}{$[\%]$} &&
           \multicolumn{2}{c}{$[10^{-4}]$} &&
           \multicolumn{2}{c}{$[10^{-2}]$} &
           \multicolumn{1}{c}{$[\!\kns]$} &
           \multicolumn{1}{c}{$[P]$}\\
         &&\multicolumn{1}{c}{(\sistd)} &&
         & \multicolumn{1}{c}{(\sistd)} &&
         & \multicolumn{1}{c}{(\sistd)} &&
         & \multicolumn{1}{c}{(\sistd)} &&
         & \multicolumn{1}{c}{(\sistd)}\\[1.0ex]\hline\\[-1.8ex]
\multicolumn{4}{l}{\textsc{Position coupled models}} &
\multicolumn{9}{l}{(illustrated with the symbol '$\circ$')}\\\hline\\[-1.8ex]
R07FU2C18  & $0.53$         & $(0.27)$ && $5.3$         & $(0.88)$&& $36$          & $(0.16)$&& $16$          & $(0.87)$ && 38  & $(2.6)$ &       & $240$ & i         & H$_{\oplus}$ \\
R07FU2C20  & $1.8$          & $(1.9)$  && $\mathbf{23}$ & $(1.7)$ && $\mathbf{65}$ & $(6.6)$ && $\mathbf{36}$ & $(3.7)$  && 54  & $(13)$  &       & $180$ & 2q        & H \\
R07FU2C25  & $3.3$          & $(2.5)$  && $33$          & $(2.5)$ && $69$          & $(11)$  && $59$          & $(9.5)$  && 110 & $(25)$  &       & $150$ & 1\~q  & H \\
R07FU4C25  & $8.2$          & $(11)$   && $36$          & $(2.4)$ && $74$          & $(14)$  && $61$          & $(12)$   && 230 & $(50)$  &       & $170$ & i         & H \\[0.5ex]
R10FU2C15  & $\mathbf{2.4}$ & $(0.53)$ && $4.9$         & $(0.32)$&& $42$          & $(1.2)$ && $12$          & $(0.30)$ && 110 & $(6.6)$ &       & $360$ & i         & H$_{\circledcirc}$ \\
R10FU2C16  & $\mathbf{11}$  & $(9.0)$  && $19$          & $(1.7)$ && $75$          & $(7.1)$ && $26$          & $(2.5)$  && 150 & $(50)$  &       & $290$ & i         & H \\
R10FU2C18  & $15$           & $(12)$   && $25$          & $(1.6)$ && $74$          & $(11)$  && $33$          & $(4.9)$  && 210 & $(59)$  &       & $250$ & 2p        & H \\
R10FU2C20  & $\mathbf{14}$  & $(12)$   && $29$          & $(1.9)$ && $\mathbf{73}$ & $(12)$  && $\mathbf{41}$ & $(7.7)$  && 240 & $(49)$  &       & $240$ & 2q    & H \\[0.5ex]
R13FU2C13  & \multicolumn{2}{c}{---}   &&               &         &&               &         &&               &          &&     &         &       &       & ---       &   \\
R13FU2C14  & $\mathbf{7.4}$ & $(4.4)$  && $7.9$         & $(1.4)$ && $48$          & $(6.0)$ && $\mathbf{11}$ & $(1.8)$  && 150 & $(14)$  &       & $450$ & i         & H$_{\circledcirc}$  \\
R13FU4C14  & $\mathbf{55}$  & $(36)$   && $14$          & $(1.4)$ && $\mathbf{66}$ & $(11)$  && $15$          & $(2.3)$  && 350 & $(85)$  &       & $400$ & i \\
R13FU2C16  & $29$           & $(19)$   && $21$          & $(1.6)$ && $71$          & $(9.1)$ && $24$          & $(3.1)$  && 270 & $(74)$  &       & $370$ & i         & H\\[0.5ex]
\multicolumn{4}{l}{\textsc{Drift models}} &
\multicolumn{9}{l}{(illustrated with the symbol '$\bullet$')}\\\hline\\[-1.8ex]
R07FU2C18  & \multicolumn{2}{c}{---}   &&               &         &&               &         &&               &          &&      &         &       &       & t \\
R07FU2C20  & $1.4$          & $(1.5)$  && $21$          & $(2.4)$ && $17$          & $(30)$  && $27$          & $(150)$  &&$63$  & $(30)$  & $13$  & $115$ & i & \dpa \\
R07FU2C25  & $1.9$          & $(2.5)$  && $34$          & $(2.9)$ && $31$          & $(29)$  && $57$          & $(280)$  &&$98$  & $(33)$  & $11$  & $87$  & i & \dpa   \\
R07FU4C25  & $5.4$          & $(8.5)$  && $33$          & $(3.0)$ && $35$          & $(40)$  && $34$          & $(99)$   &&$240$ & $(49)$  & $4.2$ & $137$ & i & \dpa   \\[0.5ex]
R10FU2C15  & $4.4$          & $(5.5)$  && $13$          & $(2.7)$ && $40$          & $(34)$  && $16$          & $(54)$   &&$57$  & $(36)$  & $9.7$ & $120$ & i & \upa$_{\bstar}$ \\
R10FU2C16  & $8.7$          & $(8.6)$  && $16$          & $(2.1)$ && $59$          & $(34)$  && $34$          & $(85)$   &&$120$ & $(46)$  & $5.6$ & $250$ & i & \dpa   \\
R10FU2C18  & $9.5$          & $(11)$   && $22$          & $(2.2)$ && $47$          & $(37)$  && $22$          & $(52)$   &&$150$ & $(54)$  & $4.0$ & $65$  & i & \dpa   \\
R10FU2C20  & $8.8$          & $(13)$   && $28$          & $(2.8)$ && $47$          & $(40)$  && $29$          & $(56)$   &&$160$ & $(42)$  & $3.8$ & $60$  & i & \dpa   \\[0.5ex]
R13FU2C13  & $15$           & $(9.7)$  && $10$          & $(0.88)$&& $47$          & $(30)$  && $7.8$         & $(12)$   &&$98$  & $(33)$  & $2.8$ & $160$ & 3\~p & $\circledast_{\bstar}$ \\
R13FU2C14  & $18$           & $(15)$   && $13$          & $(1.1)$ && $66$          & $(26)$  && $12$          & $(15)$   &&$160$ & $(45)$  & $2.5$ & $440$ & i  & \upa$_{\bstar}$ \\
R13FU4C14  & $56$           & $(44)$   && $13$          & $(1.5)$ && $54$          & $(31)$  && $12$          & $(10)$   &&$350$ & $(100)$ & $2.3$ & $170$ & i  & \\
R13FU2C16  & $23$           & $(25)$   && $20$          & $(1.8)$ && $41$          & $(39)$  && $15$          & $(24)$   &&$220$ & $(63)$  & $3.8$ & $110$ & i  & \dpa \\[1.0ex]\hline\\[-1.8ex]
\end{tabular}
\end{table*}


\subsubsection{Comparison of new PC model results with previous results}\label{sec:resultcg_old}
Before we study the drift models we comment on the agreement of the values given for the PC models in Table~\ref{tab:resultcg} (with an `H' in the last column) with the corresponding values given in {\rHoDo}. The physics of the part of the RHD3-system required to run PC models is not modified, and the result should therefore be the same (or very similar). The PC model values shown in bold face in Table~\ref{tab:resultcg} are values where the difference is $\ge\,10\%$. Comparing old and new values we find that the agreement generally is fine.

A factor that explains several of the differing values is that all new PC models are calculated for significantly longer time-intervals than the models in {\rHoDo}. Models that show different average values due to the longer time-interval include R07FU2C20, R10FU2C15, R10FU2C16 and R13FU2C14. The latter two models show large variations during parts of the evolution and the last model relaxes to a \emph{stable} pattern of variability only after several hundred piston periods. One model where the above arguments do not apply and the average values of the new models are different is R10FU2C20. For this model all average values are about ten percent larger, except for the mass loss rate which is 27\% larger. In this case it is the selection of the time-interval used in the calculation of the average properties that is important (compare with the situation of the Planck mean models in Sect.~\ref{sec:resultpg_old}).

\subsubsection{Comparison of drift models with PC models}\label{sec:resultcg_drift}
A direct comparison of the values of PC models and the corresponding values of drift models in Table~\ref{tab:resultcg} shows significant, but not dramatic changes. In this subsection we study correlations we can see by comparing values of drift and PC models.

\begin{figure*}\centering
  \caption{Ratios of four of the drift/PC model quantities given in Tables~\ref{tab:resultcg} \& \ref{tab:resultpg} plotted as a function of the respective PC model quantity. From the left the averaged quantities are: {\bf a)} the mass loss rate {\mmdot}; {\bf b)} the terminal velocity {\muinf}; {\bf c)} the degree of condensation {\mfcond}; and {\bf d)} the dust/gas density ratio {\mdrhog}. Note that all plots are logarithmic on the y-axis and of the same scale. Only models where both PC models and drift models have formed a wind are shown. The ratios of constant-opacity models are represented by open circles $\circ$, and Planck mean models by filled squares $\blacksquare$, respectively. The two constant-opacity models showing an increased mass loss rate when allowing drift are indicated in all panels with the symbol $\circledcirc$. The region enclosed by the horizontal lines, $0.9\le\,$ratio$\,\le1.1$, indicates an insignificant decrease/increase of the respective ratio. Note that each set of constant-opacity models and Planck mean models tend to form groups; excepting the two '$\circledcirc$' models.}
  \label{fig:results_ratios}
\end{figure*}

The direct comparison between PC models and drift models is aided by a plot of the ratios of the values, Fig.~\ref{fig:results_ratios}. In this subsection we only discuss the models represented by circular symbols and stars ($\circ,\oplus,\circledcirc,\bullet,\bstar$) in Figs.~\ref{fig:results_ratios}-\ref{fig:results_alfa}.

\begin{figure*}\centering
  \caption{Model average quantities as a function of the mass loss rate {\mmdot}. The panels show: {\bf a)} \& {\bf b)} the degree of condensation {\mfcond} and {\bf c)} \& {\bf d)} the optical depth of the dust {\mtaud}. The panels on the left show PC models (open symbols) and the panels on the right drift models (filled symbols). Furthermore, constant-opacity models are represented by circles, stars, and downwards pointing triangles ($\circ,\oplus,\circledcirc,\bullet,\bstar,\tridn$), and Planck mean models by squares and upwards pointing triangles ($\Box,\boxplus,\boxfd,\triup$). A key to the symbols is given just before Sect.~\ref{sec:resultcg}, and in Tables~\ref{tab:resultcg}-\ref{tab:discusspr}. Note that the PC models without a corresponding drift model (indicated by $\oplus$ \& $\boxplus$) all have low mass loss rates. The drift models are as a group better correlated with the mass loss rate in the degree of condensation ({\bf b}) than the PC models are ({\bf a}).}
\label{fig:results_mdot}
\end{figure*}

Comparing the model ratios in the panels of Fig.~\ref{fig:results_ratios} we find ratios both significantly smaller and larger than $1$. With an exception of the ratios of two models that seem to be separated from the rest of the model ratios in Figs.~\ref{fig:results_ratios}a-c (see below), the other models are grouped rather tightly. The ratio of the mass loss rates is slightly lower than 1 for the group of constant-opacity models, and the terminal velocity ratio is at a constant level close to 1. The degree of condensation is fairly constant at about $\fcond=0.7$ for PC models; the drift models show a larger scatter of values, where the individual values always are smaller than those of PC models. The two models that are separated from the group, i.e.\@ models R10FU2C15 and R13FU2C14, both share an increased drift/PC model ratio in the mass loss rate, the terminal velocity, and the dust/gas density ratio. In addition both models share a lower degree of condensation and a low terminal velocity compared to the other constant-opacity models. These two models are discussed further in Sect.~\ref{sec:discussvc}.

To illustrate correlations between the average model quantities we show Figs.~\ref{fig:results_mdot}~\&~\ref{fig:results_alfa}. Figure~\ref{fig:results_mdot} relates the degree of condensation {\mfcond} (upper panels) and the dust optical depth {\mtaud} (lower panels) to the mass loss rate. The degree of condensation measures how easily dust forms and the dust opacity shows what influence the amount of dust in the wind has on the wind structure. In the current context we mention the three models that give a higher mass loss rate in the drift models (i.e.\@ models R10FU2C15, R13FU2C13 \& R13FU2C14), indicated with stars (\bstar, drift models) and open rings ($\circledcirc$, PC models).

Figure~\ref{fig:results_mdot} shows that for the PC models there is no correlation of the degree of condensation with the mass loss rate, in addition three models (R10FU2C15, R13FU2C14 and the transition model R07FU2C18) share a significantly lower degree of condensation. The drift models, however, seem to follow a trend of increased mass loss rate with the degree of condensation (or vice versa). Both the PC models and the drift models show a correlation of the mass loss rate with the dust optical depth.

\begin{figure*}\centering
  \caption{Model average quantities as a function of {\alfa}, a quantity characterizing the strength of the radiation pressure relative to the gravitation. The panels show: {\bf a)} \& {\bf b)} the mass loss rate {\mmdot} and {\bf c)} \& {\bf d)} the terminal velocity {\muinf}. The panels on the left show PC models (open symbols) and the panels on the right drift models (filled symbols). Furthermore, constant-opacity models are represented by circles, stars, and downwards pointing triangles ($\circ,\oplus,\circledcirc,\bullet,\bstar,\tridn$), and Planck mean models by squares and upwards pointing triangles ($\Box,\boxplus,\boxfd,\triup$). A key to the symbols is given just before Sect.~\ref{sec:resultcg}, and in Tables~\ref{tab:resultcg}-\ref{tab:discusspr}. Note that the PC models without a corresponding drift model (indicated by $\oplus$ \& $\boxplus$) all are grouped in the lower-left corners in {\bf a} \& {\bf c}. The Planck mean and constant-opacity PC models are clearly more separated as groups compared to the corresponding (opacity) groups of the drift models. Also compare {\bf c} with Fig.~3 in {\rHoDo}.}\label{fig:results_alfa}
\end{figure*}

As pointed out by {\rHoDo} there should be a close correlation in dust-driven winds between the terminal velocity and the strength of the radiation pressure relative to the gravitational pull. In Fig.~\ref{fig:results_alfa} we plot the terminal velocity (lower panels) and the mass loss rate (upper panels) against the quantity {\alfa}. We adopt the same expression for {\alfa} as used by {\rHoDo} (Eq.~(1)), where it is defined at the outer boundary; {\alfa} is \emph{proportional} to the ratio of the dust radiation pressure term ($f_{\mathrm{rad,d}}$) and the gravitational force acting on the gas ($f_{\mathrm{grav,g}}$) in the (PC model) equation of motion (see Eq.~(1) in {\rSaHo}). The terminal velocity is, however, affected in the entire wind formation region and a more certain result might be given with a radially averaged {\alfa}. {\rWiLBe} calculate a similar property ($\alpha_{\mathrm{t}}$) for the innermost dust shell using the radiative pressure on both the gas and the dust. Note that {\alfa}, by definition, is not a dimensionless quantity, and therefore a value $>1\,(<1)$ does not indicate that the radiation pressure (gravitational pull) dominates.

Figure~\ref{fig:results_alfa} shows that the mass loss rate is not correlated to {\alfa}, neither for the drift models nor the PC models, also compare the scatter of models in Fig.~\ref{fig:results_alfa}a with Fig.~2 in {\rWiLBe}. A strong correlation of the velocity to {\alfa} is seen in the PC model plot of the terminal velocity in Fig.~\ref{fig:results_alfa}c, in agreement with Fig.~3 in {\rHoDo}. The agreement with Fig.~3 in {\rWiLBe} is less obvious. Because of the larger scatter of values the drift models are less well correlated in both panels; which could be an indicator of the weaker coupling between gas and dust in these models. Note that the three models that give an increased mass loss rate in drift models all are situated in the region of low {\alfa}, for both PC and drift models.

The drift velocity in Col.~12 in Table~\ref{tab:resultcg} requires a separate discussion. Since the dust velocity often shows large and dramatic variations in regions in front of shocks (see discussion in Sect.~4.2 in {\rSaHo}), it is not possible to calculate an average value for all models using the same method as for the other quantities. Instead we base this average on the longest time-interval that does not show the largest unphysical values. A small average drift velocity indicates a frequent passing of shocks (where the drift velocity is low, cf.\@ Sect.~4.2 in {\rSaHo}), and a large value correspondingly few shocks. The values in the table show a tendency towards larger drift velocities with less massive winds. We do not give a standard deviation for this quantity.

\subsection{Results of Planck mean models}\label{sec:resultpg}
The results of the Planck mean models are given in Table~\ref{tab:resultpg}. The PC models are presented in the upper part and drift models in the lower part. In contrast to constant-opacity models, Planck mean models have a lot less mass in the model domain and are more rapidly depleted of material. All PC models in this subsection are evolved for a period long enough to show a significant depletion. Consequently it is not suitable to use the same long time-intervals in the calculations of averages that we used for constant-opacity models, instead we set the upper limit of the time-interval at a time before the depletion becomes significant.

\begin{table*}
\caption{Model quantities temporally averaged at the outer boundary for Planck mean models. The numbers printed in subscript in the $T_{\mathrm{tot}}$ column indicates the upper limit in the time-interval used in the calculation of the average quantities of the corresponding model (cf\@.~Sect.~\ref{sec:resultpg_old}). The values shown in boldface for the PC models indicate that the value differ from the corresponding value given by {\rHoJoLoAr} (by $\ge\!10\%$). For the PC models an `H' in the last column indicates that the same model parameters were used in a model in {\rHoJoLoAr}. See the caption of Table~\ref{tab:resultcg} for further details.}
\label{tab:resultpg}
\begin{tabular}{lr@{\ \ }rcr@{\ \ }rcr@{\ \ }rcr@{\ \ }rcr@{\ \ }rrrrl}\hline\hline\\[-1.8ex]
   model & \multicolumn{2}{c}{\mmdot} &&
           \multicolumn{2}{c}{\muinf} &&
           \multicolumn{2}{c}{\mfcond} &&
           \multicolumn{2}{c}{\mdrhog} &&
           \multicolumn{2}{c}{\mtaud} &
           \multicolumn{1}{c}{\mvdri} &
           \multicolumn{1}{c}{$T_{\mathrm{tot}}$} &
           type \\
           \multicolumn{3}{r}{\mmdotv} &&
           \multicolumn{2}{c}{$[\!\kns]$} &&
           \multicolumn{2}{c}{$[\%]$} &&
           \multicolumn{2}{c}{$[10^{-4}]$} &&
           \multicolumn{2}{c}{$[10^{-2}]$} &
           \multicolumn{1}{c}{$[\!\kns]$} &
           \multicolumn{1}{c}{$[P]$}\\
         &&\multicolumn{1}{c}{(\sistd)} &&
         & \multicolumn{1}{c}{(\sistd)} &&
         & \multicolumn{1}{c}{(\sistd)} &&
         & \multicolumn{1}{c}{(\sistd)} &&
         & \multicolumn{1}{c}{(\sistd)}\\[1.0ex]\hline\\[-1.8ex]
\multicolumn{4}{l}{\textsc{Position coupled models}} &
\multicolumn{7}{l}{(illustrated with the symbol '$\Box$')}\\\hline\\[-1.8ex]
P07FU6C14 & $\mathbf{0.60}$& $(0.069)$ && $1.7$         & $(0.10)$ && $\mathbf{55}$ & $(0.44)$  && $\mathbf{12}$ & $(0.10)$ && $55$ & $(9.9)$ &       & $350$       & i & H$_{\boxplus}$\\
P07FU4C18 & $0.21$         & $(0.015)$ && $6.4$         & $(0.10)$ && $13$          & $(0.099)$ && $5.6$         & $(0.04)$ && $6.5$& $(0.3)$ &       & $330$       & i & H$_{\boxplus}$ \\[0.5ex]
P10FU4C14 & $0.71$         & $(0.12)$  && $2.2$         & $(0.18)$ && $\mathbf{23}$ & $(0.22)$  && $5.2$         & $(0.05)$ && $23$ & $(1.8)$ &       & $480$       & i & H$_{\boxplus}$\\
P10FU2C18 & $0.18$         & $(0.019)$ && $8.6$         & $(0.14)$ && $4.6$         & $(0.18)$  && $2.1$         & $(0.08)$ && $1.6$& $(0.1)$ &       & $_{140}430$ & i & H$_{\boxplus}$\\
P10FU4C18 & $\mathbf{1.1}$ & $(0.49)$  && $14$          & $(0.77)$ && $\mathbf{16}$ & $(2.6)$   && $7.2$         & $(1.2)$  && $7.3$& $(2.4)$ &       & $_{150}410$ & i & H\\
P10FU6C16 & $2.7$          & $(1.4)$   && $13$          & $(1.1)$  && $27$          & $(4.0)$   && $9.2$         & $(1.3)$  && $22$ & $(11)$  &       & $_{180}390$ & i \\
P10FU6C18 & $\mathbf{2.0}$ & $(0.61)$  && $16$          & $(0.48)$ && $21$          & $(1.8)$   && $9.3$         & $(0.83)$ && $15$ & $(4.9)$ &       & $_{120}390$ & i & H\\[0.5ex]
P13FU6C14 & $\mathbf{4.7}$ & $(2.3)$   && $\mathbf{10}$ & $(0.76)$ && $26$          & $(2.7)$   && $6.0$         & $(0.61)$ && $34$ & $(16)$  &       & $_{110}520$ & i & H\\
P13FU4C16 & $2.3$          & $(0.39)$  && $12$          & $(0.30)$ && $15$          & $(1.1)$   && $5.2$         & $(0.39)$ && $8.6$& $(1.7)$ &       & $_{140}440$ & i \\
P13FU6C16 & $3.9$          & $(0.51)$  && $14$          & $(0.23)$ && $21$          & $(9.6)$   && $7.3$         & $(0.33)$ && $19$ & $(2.0)$ &       & $_{80}390$  & i \\[0.5ex]
\multicolumn{4}{l}{\textsc{Drift models}} &
\multicolumn{7}{l}{(illustrated with the symbol '\boxfd')}\\\hline\\[-1.8ex]
P07FU6C14 & \multicolumn{2}{c}{---}   &&               &          &&               &           &&               &          &&       &        &       &             & --- \\
P07FU4C18 & \multicolumn{2}{c}{---}   &&               &          &&               &           &&               &          &&       &        &       &             & --- \\[0.5ex]
P10FU4C14 & \multicolumn{2}{c}{---}   &&               &          &&               &           &&               &          &&       &        &       &             & --- \\
P10FU2C18 & \multicolumn{2}{c}{---}   &&               &          &&               &           &&               &          &&       &        &       &             & t \\
P10FU4C18 & $0.82$         & $(0.12)$ && $11$          & $(0.24)$ && $12$          & $(5.3)$   && $5.4$         & $(2.9)$  && $3.3$ & $(0.5)$& $4.5$ &       $93$  & i & \dpa \\
P10FU6C16 & $1.1$          & $(1.0)$  && $8.5$         & $(1.1)$  && $19$          & $(18)$    && $7.9$         & $(13)$   && $8.7$ & $(5.6)$& $7.2$ &       $110$ & i & \dpa \\
P10FU6C18 & $2.2$          & $(2.8)$  && $22$          & $(2.3)$  && $27$          & $(34)$    && $39$          & $(210)$  && $25$  & $(10)$ & $16$  &        $65$ & 2\~q &      \\[0.5ex]
P13FU6C14 & $1.9$          & $(2.0)$  && $7.5$         & $(1.5)$  && $24$          & $(24)$    && $8.4$         & $(30)$   && $17$  & $(16)$ & $10$  &       $130$ & i & \dpa \\
P13FU4C16 & $2.5$          & $(0.69)$ && $13$          & $(0.43)$ && $23$          & $(24)$    && $11$          & $(18)$   && $8.7$ & $(1.5)$& $4.9$ &       $130$ & 2\~q &      \\
P13FU6C16 & $4.1$          & $(4.0)$  && $14$          & $(1.1)$  && $19$          & $(21)$    && $11$          & $(39)$   && $16$  & $(4.1)$& $4.0$ &       $130$ & i\\[1.0ex]\hline\\[-1.8ex]
\end{tabular}
\end{table*}



\subsubsection{Comparison of new PC model results with previous results}\label{sec:resultpg_old}
Like for the constant-opacity models in Sect.~\ref{sec:resultcg_old} we comment on the agreement of the values given for the PC models (indicated with an `H' in the last column) in Table~\ref{tab:resultpg} with the corresponding values given in {\rHoJoLoAr}. The result should be the same (or very similar). The PC model values shown in bold face in Table~\ref{tab:resultpg} are values where the difference is $\ge\!10\%$. 

The length of the time-interval used in the calculation of average quantities is a factor that can explain several of the different values in Sect.~\ref{sec:resultcg_old}. Because of the fast depletion of the model domain in the Planck mean models this it is not the cause of discrepancy here. Instead we point out that it is the detailed selection of the time-interval used in the calculation of the averages that is crucial. We select this upper time-interval limit for models showing depletion to be where the \emph{average} decrease in the mass loss rate is insignificant. This point can be very difficult to define in winds of irregular variability and it is therefore wise to choose a shorter period. The approximate time (in piston periods) printed in subscript in the last but two columns in Table~\ref{tab:resultpg} indicates this upper limit.

All models that differ by $\ge\!10\%$ in Table~\ref{tab:resultpg}, except P07FU6C14, are found to differ due to the selection of the time interval. Model P07FU6C14 evolves very slowly to a stable variability pattern, and it could be that we measure the quantities during different states of the wind (using different time-intervals).

\subsubsection{Comparison of drift models with PC models}\label{sec:resultpg_drift}
Due to the overall lower density in the Planck mean models these models form a lot less dust than the constant-opacity models do. The results of this are for example a lower average degree of condensation and a lower average mass loss rate in the former type of models \citep[cf.\@ {\rHoJoLoAr} and][]{HeWiSe:00}. A first difference seen in the drift models of Table~\ref{tab:resultpg} is that only three of the sets of {\rHoJoLoAr}-PC model parameters are found to form a drift model wind. None of the PC models with a mass loss rate $<\!1.0\times10^{-6}\,\mdotu$ successfully forms a drift model wind. With fewer drift models the statistics is on the one hand less accurate and correlations become less certain. On the other hand, with the current selection of model parameters, we get stronger \emph{conditions} on the limit for wind formation in drift models (cf.\@ Sect.~\ref{sec:discussv}).

The results of all Planck mean models are plotted together with the constant-opacity models in Figs.~\ref{fig:results_ratios}-\ref{fig:results_alfa}. In this subsection we study models represented by the rectangular symbols ($\Box,\boxplus,\boxfd$) and compare with the relations of the constant-opacity models. We again find both increased and decreased values in Fig.~\ref{fig:results_ratios}. There is, however, no model (or group of models) showing a significantly increased mass loss rate (like two constant-opacity models do). While the values of the PC models group clearly around $13\kms$ in Fig.~\ref{fig:results_ratios}b, around $0.2 $ in Fig.~\ref{fig:results_ratios}c, and around $0.8\times10^{-3}$ in Fig.~\ref{fig:results_ratios}d, there is a larger scatter in the values of the drift models. There is no evident correspondence with the scatter of the constant-opacity models. Like for the constant-opacity models the (gray) triangles (of the Planck mean models in Table~\ref{tab:discusspr}, Sect.~\ref{sec:discussp}) show a distribution similar to the rest of the drift models.

In Fig.~\ref{fig:results_mdot} (and Fig.~\ref{fig:results_alfa}) it seems that the Planck mean PC models are scattered over a larger region than the constant-opacity PC models are. However, if those models that do not give a corresponding drift model are excluded (indicated by $\boxplus$), six tightly grouped models are left. For the drift models it appears that the Planck mean models correlate well with the constant-opacity models, in particular in the dust optical depth (Fig.~\ref{fig:results_mdot}d). The drift models in Fig.~\ref{fig:results_alfa} share the properties of -- and show a distribution similar to that of -- the constant-opacity models. Excepting one model, P10FU6C18, the Planck mean models compared to the constant-opacity models, on average share a lower value of {\alfa} (this is caused by a lower dust/gas density ratio, Fig.~\ref{fig:results_ratios}d).

Note that all PC models with a terminal velocity below $10\kms$ either lack a corresponding drift model, or show significantly different properties when drift is included. The average drift velocities in the six drift models are all fairly low, and two models are even multi-periodic during parts of the calculated time-interval (cf.\@ the discussion in Sect.~\ref{sec:discussp}). In addition to the variability associated with the shell structure of the winds, the dust velocity shows small-amplitude variations at the outer boundary for all drift models in Table~\ref{tab:resultpg} (see Sect.~\ref{sec:discussv}).

\section{Discussion}\label{sec:discuss}


\begin{figure*}\centering
  \caption{Radial structure of the inner parts of the wind for the drift (solid line) and PC (dash-dot-dotted line) models R10FU2C18. The panels show: {\bf a)} the gas velocity $u$; {\bf b)} the average grain radius $\langle r_{\mathrm{d}}\rangle$; {\bf c)} the gas density $\rho$; {\bf d)} the degree of condensation {\fcond}; {\bf e)} the drift velocity {\vdri}; and {\bf f)} the grain/gas number density ratio $n_{\mathrm{d}}/n_{\mathrm{g}}$. The accumulation of the dust to regions behind shocks in drift models is seen in e.g.\@ the dust/gas number density ratio ({\bf f}) and the degree of condensation ({\bf d}). Note the larger variations in the dust quantities of the drift model (right panels).}
\label{fig:discussw}
\end{figure*}

In the current models drift does not change the fundamental processes behind the wind formation mechanism, but is still able to alter the conditions for dust formation. In wind models allowing drift the dust tends to accumulate to dense regions behind shocks. The reason for this relocation is that the gas-dust collisional interaction is weak in inter-shock regions and the dust, accelerated by the radiation pressure, drifts with respect to the gas until the gas-dust interaction is strong enough to slow the drift down. The dense regions behind shocks provide such conditions (cf.\@ {\rSaHo}).

Figure~\ref{fig:discussw} shows the spatial structure of the constant-opacity drift and PC models R10FU2C18. The quantities are selected to illustrate the difference in the spatial wind structures of the two models. The accumulation of dust to the regions behind shocks is evident in the degree of condensation and the dust/gas number density ratio; the dust density is several orders of magnitude lower in the inter-shock region, where the drift velocity on the contrary is higher. The shock/inter-shock gas density ratio is much lower than that of the dust. Although the dust densities of the two models differ, the gas structures appear similar in the gas velocity and the gas density.

Figures~\ref{fig:discussw}b,f show that both the average grain radius and the grain abundance vary more strongly in the drift model. The average grain radius is typically below $10^{-5}\,$cm, justifying the use of the small particle limit in the calculation of the dust opacities. The high values in the innermost peak in the plot of the average grain radius, at $2\,R_*$ are irrelevant since the degree of condensation, and therefore the dust opacity, at this location is negligible (compare Fig.~\ref{fig:discussw}b with Fig.~\ref{fig:discussw}d).

In most cases of our study the consequences of the accumulation of dust to shocks tends to be (on average) narrower regions of efficient nucleation and grain growth. This in turn leads to smaller total amounts of formed dust, a less efficient wind formation, and lower mass loss rates compared to winds formed in PC models. However, this is not necessarily the case if the dust formation in a certain PC model is inefficient. In a few cases the accumulation of dust to narrow regions (or shells) in drift models seems to provide more favorable conditions for the wind formation with a resulting increased mass loss rate.

In this section we study four topics of the wind formation where the influence of drift is of a primary concern. We first study the influence of the piston period in the next subsection. In Sect.~\ref{sec:discussv} we measure and compare the variability of the wind models.
The comparison with a derived mass loss rate fit formula is discussed in Sect.~\ref{sec:discussm}, both for the PC and the drift models and for constant-opacity models vs.\@ Planck mean models.
Finally we comment on the importance of the heating by drift for the energy balance in drift models in Sect.~\ref{sec:discussh}.

\subsection{Influence of the piston period}\label{sec:discussp}
We believe it is important to follow up the study regarding the influence of the piston period on the wind structure carried out by {\rWiLBe} (Sect.~3.5.6) and {\rHoDo} for the case of drift models. In particular regarding Planck mean models, for which there is no similar study. All models discussed this far are calculated with a P-L relation derived for M-type Miras by \citet[we refer to these models as F-models]{FeGlWhCa:89}. A more recent relation derived for C-rich Miras, derived by \citet[G-models]{GrWh:96}, reveal longer periods (see Sect.~\ref{sec:physpuls}). The results of seven G-models are presented in Table~\ref{tab:discusspr} together with the results of the respective F-models for comparison. The wind properties that differ by more than $10\%$ from the corresponding F-model are indicated in boldface. The piston periods of the (F- and) G-models are given in Table~\ref{tab:detmodel}.

\begin{table*}
\caption{Results of drift models using two different P-L relations. F-Models are indicated with an ``F'' in the model name (first column), and G-models with a ``G'', cf\@. Sects.~\ref{sec:modelp} \& \ref{sec:discussp}. The symbols in the last column indicate whether the mass loss rates of the G-models have an increased/decreased (\upa,\dpa) value compared to the corresponding F-models. The values shown in boldface for the G-models indicate that the value differ from the corresponding F-model value (by $\ge\!10\%$). The symbol printed in subscript for the F-models in the last column indicates how the respective model is illustrated in Figs.~\ref{fig:results_mdot}, \ref{fig:results_alfa} \& \ref{fig:discuss_dev}. See the caption of Table~\ref{tab:resultcg} for further details.}
\label{tab:discusspr}
\begin{tabular}{lr@{\ \ }rcr@{\ \ }rcr@{\ \ }rcr@{\ \ }rcr@{\ \ }rrrrl}\hline\hline\\[-1.8ex]
   model & \multicolumn{2}{c}{\mmdot} &&
           \multicolumn{2}{c}{\muinf} &&
           \multicolumn{2}{c}{\mfcond} &&
           \multicolumn{2}{c}{\mdrhog} &&
           \multicolumn{2}{c}{\mtaud} &
           \multicolumn{1}{c}{\mvdri} &
           \multicolumn{1}{c}{$T_{\mathrm{tot}}$} &
           type \\
           \multicolumn{3}{r}{\mmdotv} &&
           \multicolumn{2}{c}{$[\!\kns]$} &&
           \multicolumn{2}{c}{$[\%]$} &&
           \multicolumn{2}{c}{$[10^{-4}]$} &&
           \multicolumn{2}{c}{$[10^{-2}]$} &
           \multicolumn{1}{c}{$[\!\kns]$} &
           \multicolumn{1}{c}{$[P]$}\\
         &&\multicolumn{1}{c}{(\sistd)} &&
         & \multicolumn{1}{c}{(\sistd)} &&
         & \multicolumn{1}{c}{(\sistd)} &&
         & \multicolumn{1}{c}{(\sistd)} &&
         & \multicolumn{1}{c}{(\sistd)}\\[1.0ex]\hline\\[-1.8ex]
\multicolumn{4}{l}{\textsc{constant-opacity drift models}} &
\multicolumn{9}{l}{(G-models are illustrated with the symbol '\tridn')}\\\hline\\[-1.8ex]
R10FU2C15 & $4.4$           & $(5.5)$  && $13$          & $(2.7)$ && $40$          & $(34)$  && $16$          & $(54)$  && $57$          & $(36)$  & $9.7$ & $120$ & i & $|_{\bstar}$\\
R10GU2C15 & $\mathbf{5.2}$  & $(4.1)$  && $\mathbf{16}$ & $(1.2)$ && $\mathbf{49}$ & $(20)$  && $17$          & $(60)$  && $\mathbf{80}$ & $(28)$  & $\mathbf{4.4}$ &  $75$ & 2p & \upa\\[0.5ex]
R10FU2C18 & $9.5$           & $(11)$   && $22$          & $(2.2)$ && $47$          & $(37)$  && $22$          & $(52)$  && $150$         & $(54)$  & $4.0$ &  $65$ & i & $|_{\bullet}$\\
R10GU2C18 & $\mathbf{6.8}$  & $(7.9)$  && $24$          & $(2.1)$ && $\mathbf{25}$ & $(37)$  && $22$          & $(80)$  && $\mathbf{120}$& $(38)$  & $\mathbf{7.1}$ &  $65$ & 2\~q & \dpa\\[0.5ex]
R13FU2C14 & $18$            & $(15)$   && $13$          & $(1.1)$ && $66$          & $(26)$  && $12$          & $(15)$  && $160$         & $(45)$  & $2.5$ & $220$ & i & $|_{\bstar}$\\
R13GU2C14 & $\mathbf{22}$   & $(21)$   && $13$          & $(1.4)$ && $68$          & $(20)$  && $12$          & $(7.5)$ && $\mathbf{190}$& $(53)$  & $2.5$ & $220$ & i & \upa\\[0.5ex]
R13FU2C16 & $23$            & $(25)$   && $20$          & $(1.8)$ && $41$          & $(39)$  && $15$          & $(24)$  && $220$         & $(63)$  & $3.8$ & $110$ & i & $|_{\bullet}$\\
R13GU2C16 & $\mathbf{17}$   & $(20)$   && $20$          & $(2.0)$ && $\mathbf{47}$ & $(38)$  && $\mathbf{17}$ & $(79)$  && $\mathbf{190}$& $(47)$  & $\mathbf{4.2}$ & $85$  & i & \dpa\\[0.5ex]
\multicolumn{4}{l}{\textsc{Planck mean drift models}} &
\multicolumn{9}{l}{(G-models are illustrated with the symbol '\triup')}\\\hline\\[-1.8ex]
P10FU6C16 & $1.1$           & $(1.0)$  && $8.5$         & $(1.1)$ && $19$          & $(18)$  && $7.9$         & $(13)$  && $8.7$         & $(5.6)$ & $7.2$ & $110$ & i & $|_{\boxfd}$\\  
P10GU6C16 & $\mathbf{2.7}$  & $(2.8)$  && $\mathbf{14}$ & $(1.4)$ && $\mathbf{30}$ & $(30)$  && $\mathbf{17}$ & $(66)$  && $\mathbf{27}$ & $(15)$  & $\mathbf{6.0}$ & $69$  & i & \upa\\[0.5ex]
P13FU6C14 & $1.9$           & $(2.0)$  && $7.5$         & $(1.5)$ && $24$          & $(24)$  && $8.4$         & $(30)$  && $17$          & $(16)$  & $10$ & $130$ & i & $|_{\boxfd}$\\
P13GU6C14 & $\mathbf{4.9}$  & $(3.0)$  && $\mathbf{9.1}$& $(0.94)$&& $\mathbf{21}$ & $(21)$  && $\mathbf{6.3}$& $(34)$  && $\mathbf{21}$ & $(9.4)$ & $\mathbf{4.7}$ & $110$& i & \upa\\[0.5ex]
P13FU6C16 & $4.1$           & $(4.0)$  && $14$          & $(1.1)$ && $19$          & $(21)$  && $11$          & $(39)$  && $16$          & $(4.1)$ & $4.0$ & $130$ & i & $|_{\boxfd}$\\   
P13GU6C16 & $\mathbf{5.6}$  & $(5.8)$  && $\mathbf{17}$ & $(1.5)$ && $\mathbf{11}$ & $(24)$  && $\mathbf{15}$ & $(90)$  && $\mathbf{35}$ & $(20)$  & $\mathbf{4.9}$ & $67$  & 2\~q & \upa\\[1.0ex]\hline\\[-1.8ex]
\end{tabular}
\end{table*}



In the \emph{constant-opacity models} (upper half in Table~\ref{tab:discusspr}) we note that the values of the G-models primarily differ in the mass loss rate and the degree of condensation. Quantitatively the changes are of the same order of magnitude as found by {\rHoDo}. The exception is model R10GU2C18 where the degree of condensation is about half of that of R10FU2C18. Two G-models, R10GU2C18 and R10GU2C15 also show multi-periodic variations where the F-models do not. These two models and the F-model R13FU2C13 are the only constant-opacity drift models in this study that are multi-periodic. In the latter model a dust shell forms every third piston period.

\emph{Planck mean models} (lower half in Table~\ref{tab:discusspr}) are more sensitive to the period than the constant-opacity models are. The values of all three G-models differ significantly compared to the corresponding F-models. One model, P13GU6C16 is even multi-periodic during parts of the evolution. This model and the two F-models P10FU6C18 \& P13FU4C16 are the only three Planck mean models (among both drift and PC models) where we find multi-periodicity. The changes show both increasing and decreasing values in all quantities. Additionally in the mass loss rate, the terminal velocity, and the dust optical depth all G-models show higher values.

We conclude that while constant-opacity models are only little affected by the piston period in some models (in accordance with the results of {\rWiLBe} and {\rHoDo}), Planck mean models are sensitive to the period to a larger extent with significant changes in all wind properties.

\subsection{The importance of variability in the wind}\label{sec:discussv}
The temporal evolution at the outer boundary is shown for the drift and PC models R13FU4C14 (constant-opacity models, left panels) and P13FU6C14 (Planck mean models, right panels) in Fig.~\ref{fig:discuss_var}. All corresponding panels on the left and the right are drawn at the same scale on the axes for easy comparison. The same model parameters are used in both the Planck mean models and the constant-opacity models, with the exception of the piston amplitude which is higher in the two Planck mean models. The higher value is used as a partial compensation for the lower densities in Planck mean models. The result of the less efficient dust formation in the Planck mean model P13FU6C14 is clearly evident through all panels in the figure.

The large variability of the drift models make a direct comparison with the PC models difficult in terms of which wind has larger average values. Although the variations in the terminal velocity and the mass loss rate seem to be of the same magnitude and time scale for the models R13FU4C14, they differ in both the PC and drift models P13FU6C14. Note that the variational patterns in the dust quantities of the drift models (lowermost four panels for both pairs of models in Fig.~\ref{fig:discuss_var}) differ significantly from the corresponding patterns of the PC models. For the two drift models shown in this figure less dust is formed, the grains are smaller, and the mass loss rate is decreased in comparison with the values of the respective PC model.

In the lowermost panels in Fig.~\ref{fig:discuss_var} we see periodic small-amplitude variations superposed on the drift velocity. These variations do not appear in the gas velocity and last for the entire calculated time-interval. They are larger in the less dense inter-shock regions. In addition, compared to the constant-opacity model, they are more pronounced in the less massive wind of the Planck mean model. The period of these variations is for both models equal to the piston period. By the strong coupling to the radiation field the dust senses the variations in the luminosity caused by the radial oscillations of the piston. The small-amplitude variations remain in the dust velocity as the collisional coupling between the gas and the dust is not strong enough to flatten them out. These variations ought to be present in all models, but since the long-term variations in the drift velocity are larger in several models they are not always obvious. Corresponding modulations are visible in all Planck mean models except the multi-periodic model P10FU6C18. In addition they appear in three constant-opacity models: R10GU2C15, R13FU4C14 and R13FU2C16.

\begin{figure*}\centering
  \caption{The temporal evolution at the outer boundary. The panels on the l.h.s.\@ show the constant-opacity drift and PC models R13FU4C14 (solid and dash-dot-dotted lines, respectively), and the panels on the r.h.s.\@ the Planck mean drift and PC models P13FU6C14 (solid and dash-dot-dotted lines, respectively). With the exception of the piston amplitude {\deltaup}, which is larger in the Planck mean models, all other model parameters are the same. The panels show (from the top): the terminal velocity $u_{\infty}$; the mass loss rate $\dot{M}$; the degree of condensation $f_{\mathrm{cond}}$; the dust/gas density ratio {\drhog}; the mean grain radius $\langle r_{\mathrm{d}}\rangle$; and the drift velocity $v_{\mathrm{D}}$. Note the much larger variations present in both drift models compared to the PC models, also compare with Fig.~\ref{fig:discuss_dev}.}
\label{fig:discuss_var}
\end{figure*}

\begin{figure*}\centering
  \caption{Temporal variations of average properties ($q$) measured by the fluctuation amplitude $r=\sistd/q$ (Sect.~\ref{sec:discussv}). From the left the figure shows: {\bf a)} the mass loss rate {\mmdot}; {\bf b)} the terminal velocity {\muinf}; {\bf c)} the degree of condensation {\mfcond}; and {\bf d)} the dust optical depth {\mtaud}. All plots are drawn to the same scale. Filled symbols represent drift models (lower panels) and open symbols PC models (upper panels). constant-opacity models are represented by circles, stars, and downwards pointing triangles ($\circ,\oplus,\circledcirc,\bullet,\bstar,\tridn$), and Planck mean models by squares and upwards pointing triangles ($\Box,\boxplus,\boxfd,\triup$). Furthermore, PC models without a corresponding drift model for the same set of model parameters are indicated with the symbol $\oplus$ or $\boxplus$. Models for which the drift models have a larger average mass loss rate are indicated with open rings ($\circledcirc$; PC models), and stars ($\star$; drift models). Triangles represent G-models (cf.\@ Sect.~\ref{sec:discussp}). Note the distinctly smaller scattering and larger variations of the values of the drift models compared to the values of the PC models.}\label{fig:discuss_dev}
\end{figure*}

To quantify the variability of the model properties we have calculated an (absolute) standard deviation ({\sistd}) for each average outflow quantity (except the drift velocity, see Sect.~\ref{sec:resultcg_drift}) given in Tables~\ref{tab:resultcg}-\ref{tab:discusspr}. However, a relative fluctuation amplitude is preferred in comparisons between different models, see below. The relative \emph{fluctuation amplitude} is simply the standard deviation divided by the average value ($r=\sistd/q$, i.e.\@ a ``relative error''). A large variability, typical of a time-dependent outflow, is represented by a large fluctuation amplitude. And a small variability, typical of a ``stationary'' outflow, by a small fluctuation amplitude. In a similar argument {\rWiLBe} used the standard deviation to quantify the stability of the wind.

In the time sequences shown in Fig.~\ref{fig:discuss_var} we saw a tendency of a larger variability in the two drift models, compared to the corresponding PC models. The fluctuation amplitude is shown in Fig.~\ref{fig:discuss_dev} as a function of four of the average properties of Tables~\ref{tab:resultcg}-\ref{tab:discusspr}. With Fig.~\ref{fig:discuss_dev} we see that the variations generally are slightly larger to significantly larger in the drift models of most drift/PC model pairs.

By marking the PC models that lack a corresponding drift model (with the symbols $\oplus$ \& $\boxplus$) we find that these models, compared to the other models, either show smaller variations or lower average values. Apparently they share several properties. For instance, all PC models in this group share a terminal velocity $<\,10\kms$, no other PC model does. Furthermore, these models tend to have a mass loss rate in the lower end of the range. We point out that we do not attempt to model winds where we initially are certain that there is no drift model wind. The statistics of the subset of PC models without corresponding drift models is therefore not sufficient for generalized quantitative conclusions in this respect.

We first study the constant-opacity models in the following subsection and then Planck mean models in Sect.~\ref{sec:discussvp}.

\subsubsection{Variability of constant-opacity models}\label{sec:discussvc}
In this subsection we discuss constant-opacity models shown with circular symbols, stars, and downwards pointing black triangles ($\circ,\oplus,\circledcirc,\bullet,\bstar,\tridn$) in Fig.~\ref{fig:discuss_dev}. The basic properties of the models are studied in Sect.~\ref{sec:resultcg_drift}.

The variability of the drift models are for most cases about the same or larger than the variability in the PC models in Fig.~\ref{fig:discuss_dev}. Two PC models show variations in the terminal velocity that are larger than those of the bulk of the drift models. These models, R07FU2C18 ($\oplus$) and R13FU2C14 ($\circledcirc$), together with R10FU2C15 ($\circledcirc$) all have a lower terminal velocity and show a lower variability in {\taud} compared to the rest of the PC models. Moreover, the latter two models, showing an increased mass loss rate in the corresponding drift models, both have a degree of condensation below $\fcond\simeq0.5$. Model R07FU2C18 barely forms a wind using drift (indicated by a `t' in Table~\ref{tab:resultcg}). In contrast the two corresponding drift models R10FU2C15 and R13FU2C14 correlate well in the variability with the other drift models.

To understand exactly what model parameters result in drift models with higher mass loss rates we would need a large sample of models that do not show any corresponding drift model. It would with this sample of models be possible to separate the PC models without drift model winds from the corresponding models giving a higher mass loss rate in drift models. These latter models can in turn, using the arguments in the previous paragraph, be separated from the rest of the models that give a similar or lower mass loss rate in the drift models.

In this context we mention the drift model R10FU2C13, which lacks a corresponding PC model wind. One attempt to form winds with a lower carbon/oxygen ratio, $\CtoO=1.25$, in a drift model with the otherwise same parameters as model R13FU2C13 failed. Similarly a second attempt with $\CtoO=1.45$ in a (PC \& drift) model with the otherwise same parameters as model R10FU2C15 also failed. The properties of the model with the closest model parameters, with a higher carbon/oxygen ratio, R13FU2C14 shows that the added freedom in the drift model is crucial in forming a wind with these parameters.

\subsubsection{Variability of Planck mean models}\label{sec:discussvp}
In this subsection we discuss Planck mean models, shown with rectangular symbols and upwards pointing triangles ($\Box,\boxplus,\boxfd,\triup$) in Fig.~\ref{fig:discuss_dev}. The basic properties of these models are studied in Sect.~\ref{sec:resultpg_drift}. A comparison of the distribution of model values in the figure between Planck mean models and constant-opacity models show that the scatter is larger for the former type of models, both for PC and drift models.

All Planck mean PC models that do not have a corresponding drift model ($\boxplus$) show a terminal velocity $\muinf\lesssim10\kms$. In addition the same four models show low mass loss rates and low variations in all quantities. The variations are particularly low in the degree of condensation where no other but these four Planck mean models show a fluctuation amplitude $r<0.05$. This indicates the importance of a time-dependent wind structure for forming drift models. We conclude that Planck mean PC models of a low terminal velocity and low variations in the degree of condensation are the models that are least likely to form a wind in a drift model.

\subsection{Comparison of the mass loss rates with a mass loss rate fit formula}\label{sec:discussm}
The calculation of wind models is a time consuming process. And it may in some cases be preferable if the calculation of the mass loss rate can be replaced by a function with a small number of parameters. \citet{WaScWi.:02} derived a mass loss rate fit formula based on the wind models in {\rWiLBe}, that are similar to our constant-opacity PC models, and the P-L relation of \citet{GrWh:96} (see Sect.~\ref{sec:physpuls}). Their expression allows the calculation of the mass loss rate using only four stellar parameters (the pulsation period $P$ is below indirectly included in the $L$-term through the P-L relation),
\begin{eqnarray}
\log\dot{M}_{\mathrm{fit}}&=&-4.52-6.81\log(T_{\mathrm{eff}}/2600\,\mbox{K})\nonumber\\&&+2.47\log(L/10^4L_{\sun})-1.95\log(M/M_{\sun})\,.
\label{eq:discussm}
\end{eqnarray}
The units of $\dot{M}_{\mathrm{fit}}$ are {\mdotu}. Our models are well within the range of stellar parameters covered by Eq.~(\ref{eq:discussm}). In this subsection we discuss how the output of our models compares to this expression.

\begin{figure}
\caption{Ratios of the average mass loss rates $\mmdot$ to the mass loss fit $\dot{M}_{\mathrm{fit}}$ given in Eq.~(\ref{eq:discussm}) as a function of: {\bf a)} \& {\bf c)}  the effective temperature $T_{\mathrm{eff}}$; and {\bf b)} \& {\bf d)} the stellar luminosity $L$. The upper panels show PC model ratios and the lower panels additionally drift model ratios (for comparison). The plot symbols are the same as in Fig.~\ref{fig:discuss_dev}. The horizontal lines indicate the mean ratio for each type of model wind. This figure (specifically the shaded region) should be compared with Figs.~1b,c in \citet{WaScWi.:02}. The Planck mean models are drawn in light gray symbols in the upper panels to emphasize that Eq.~(\ref{eq:discussm}) is derived for constant-opacity PC models; cf.\@ Sect.~\ref{sec:discussm}.}
\label{fig:discussm}
\end{figure}

The mass loss rate ratio $\mmdot/\dot{M}_{\mathrm{fit}}$ is illustrated for the models presented in this article in Fig.~\ref{fig:discussm}. A comparison of the constant-opacity PC models in this figure (illustrated by open circular symbols ($\circ,\oplus,\circledcirc$) in the upper panels) with Figs.~1b,c in \citet{WaScWi.:02} reveals systematically lower mass loss rates in our models. This discrepancy is most likely caused by a smaller levitation in our models. The models used by \citeauthor{WaScWi.:02} have a piston velocity amplitude $\deltaup=5\kms$ compared to 2\kms used in most of our constant-opacity models. The two pairs of models with 2 \& {4\kms} but otherwise identical parameters (see Table~\ref{tab:resultcg}) demonstrate the influence of {\deltaup} -- and consequently the levitation -- on the mass loss rate.

The densities in the Planck mean models are much lower compared to the corresponding values in the constant-opacity models. And the resulting mass loss rates are much lower (Sect.~\ref{sec:resultpg} \& {\rHoJoLoAr}; illustrated by squares ($\Box,\boxplus$) in Fig.~\ref{fig:discussm}). The mass loss rate also differs when drift is allowed (Sects.~\ref{sec:resultcg_drift}, \ref{sec:resultpg_drift}; lower panels in Fig.~\ref{fig:discussm}), but the change is less obvious in the figure above. The mass loss rate fit in its present form is inapplicable to these winds as it does not account for these two effects. However, the computational effort needed to calculate a new relation is hardly justified since future further physically improved models possibly quickly will invalidate such a relation.

\subsection{The influence of heating by drift}\label{sec:discussh}
The importance of heating by drift (\qdrift) for the wind structure was studied by {\rKrGaSe}, who found the heating term (cf.\@ Sect.~\ref{sec:physgdin}) to be an important factor in the wind energy balance in the context of the stationary winds they modeled. They referred to the term as $q_{\mathrm{fric}}$. In the models of the first article in this series, {\rSaHo}, we did not include this term as its inclusion is connected to several numerical difficulties. For comparison with those models it is not included in the models discussed in Sect.~\ref{sec:nonpulsa} in this article, but it is included in all other models.

Places in the wind where the heating by drift could be expected to be important are in the dense regions behind shocks where the drag force is strong. However, the drift velocity is simultaneously low at these locations, and it is therefore difficult to make a simple estimate of the importance of this heating term compared to the other terms in the energy equation. The nonlinear character of the terms in the system of equations describing the wind in connection with a highly variable structure, makes it difficult to predict when and where in the wind {\qdrift} is important.

\begin{figure}
\caption{Radial structure of the Planck mean model P10FU6C18 (solid line) and the constant-opacity model R10FU2C18 (dash-dot-dotted line). The shown quantities are: {\bf a)} the gas velocity; {\bf b)} the gas density; {\bf c)} the drift velocity; {\bf d)} the drag force; {\bf e)} \& {\bf f)} the energy ratios (see Sect.~\ref{sec:discussh}). This figure illustrates the importance of the heating by drift ({\qdrift}) for the energy balance of the respective model. The heating by drift is negligible in most parts of the Planck mean model (lowermost two panels). On the contrary it is comparable in magnitude to the other energy source terms in the constant-opacity model. Even so we do not find any significant effect on the wind structures caused by this term, cf.\@ Sect.~\ref{sec:discussh}.}
\label{fig:qdrift}
\end{figure}

With Fig.~\ref{fig:qdrift} we illustrate the complex influence of the heating by drift. The radial plot shows the Planck mean model P10FU6C18 (solid line) and the constant-opacity model R10FU2C18 (dash-dot-dotted line). The uppermost two panels in Fig.~\ref{fig:qdrift} depicting the gas velocity $u$ and the gas density $\rho$ show the locations of the shocks. The middle panels depicting the drift velocity {\vdri} and the drag force {\fdrag} are plotted with the same scale and illustrate the importance of the drag force in relation to the drift velocity in the {\qdrift} term. The lowermost two panels show the relative strength of the heating by drift {\qdrift} compared to the term describing the work done on the gas by the velocity field ($P\nabla\!\cdot\!u$, Fig.~\ref{fig:qdrift}e), and the energy exchange with the radiation field ($4\pi\rho\kappa_{\mathrm{g}}(J-S_{\mathrm{g}})$, Fig.~\ref{fig:qdrift}f; where $S_{\mathrm{g}}$ is the gas source function) respectively. In both these panels a ratio less than one indicates that the heating by drift is smaller than the respective heating term. Note the logarithmic scale in both these panels.

The spatial structures in each model vary significantly with time and the presented instants do not represent all possible states of the plotted quantities. They do, however, give an impression of the typical features and range of values. Figs.~\ref{fig:qdrift}c,d show that the variation of the drag force is more important to the {\qdrift} term than the drift velocity is. The lowermost two panels show that the same term has a negligible influence on the energy balance in the Planck mean model; the energy ratios are only close to one in the shocked regions. The situation is obviously different in the denser constant-opacity model, where the heating by drift seems to be relevant for larger zones of the model. Nevertheless, regarding the average wind properties we do not find any measurable ``significant'' differences between models calculated with and without {\qdrift}, neither in any of the Planck mean models nor in any of the constant-opacity models presented in the tables in Sect.~\ref{sec:results}.

We conclude that although the heating by drift seems to be of a minor importance for the models we have studied here, it can (possibly) not safely be neglected in general. The term {\qdrift} should therefore be included in the calculations.

\section{Conclusions}\label{sec:conclus}
In this paper we have studied the effects of drift between dust and gas in time-dependent wind models of cool long-period variables. The drift and its consequences for the wind structure have been discussed for several decades. However, until recently these discussions were almost exclusively based on stationary wind models. These models, and the associated structures, showing a smooth distribution of gas and dust may be misleading when trying to estimate the effects of drift, as studies using time-dependent models have demonstrated within the last few years. In addition, among the existing dynamical wind models only a few actually describe the formation of the stellar wind using a self-consistent description of the gas, the dust and the radiation field. In this article we have presented models using such a description \citep[i.e.\@ the approach introduced by][{\rSaHo}]{SaHo:03}.

With this study we wanted to find out how the wind properties, and the variability of the wind structure, change as a consequence of including drift in the calculations. Compared to {\rSaHo} we now include the effects of stellar pulsations (simulated by variable inner boundary conditions; cf.\@ Sect.~\ref{sec:physpuls}) and the heating of the gas caused by collisions with drifting dust grains (cf.\@ Sect.~\ref{sec:physgdin}). Concerning the interaction between the gas and the radiation field, we have studied models based on two different representations of the gas opacity adopted in the literature. On the one hand we have calculated models assuming a constant value of the mass absorption coefficient (constant-opacity models); an approach widely used in time-dependent wind models. We have also, on the other hand, calculated models based on Planck mean absorption coefficients, based on molecular data (Planck mean models). This approach results in more realistic pressure-temperature conditions in the photospheric layers, and lower densities in general (cf.\@ Sect.~\ref{sec:gasradin}).

We have selected the sets of model parameters from a sample where we expected that drift could be important to the wind structure. In the selection process we used stellar parameter relations derived from observations and evolution models to constrain the number of independent parameters -- thereby more clearly separating the effects of drift. Spatially our models cover a region ranging from below the stellar photosphere out to about 25 stellar radii, enclosing the zones of dust formation.

We have compared drift models with the corresponding position coupled (PC; non-drift) models, and have found that the general changes in the spatial structure and variability in {\rSaHo}, caused by drift, are confirmed in the larger set of models presented in this article. When drift is allowed dust tends to accumulate to the dense regions behind shocks. The gas-dust collisional interaction in these regions is comparatively strong and the drift velocity conversely low, the regions between the shocks are correspondingly depleted of dust. The consequences of the spatial relocation of the dust are larger temporal variations of dust-related quantities (at fixed locations in the wind). Gas-related quantities are also affected in this process, but to a much lower degree.

Several PC models in this and earlier studies show a \mbox{(multi-)}periodic variability in for instance the mass loss rate and the terminal velocity, whereas most drift models show irregular variations. However, we have found a couple of drift/PC model pairs where the drift model is (quasi) periodic, and the PC model is not. In particular, we mention that three of these drift models are Planck mean models. No periodic Planck mean PC models have been found here.

Considering the changes between drift models and PC models -- in the context of temporally averaged quantities -- we have found significant, but not dramatic, differences for most models. There is, for instance, a tendency towards reduced mass loss rates and a larger scatter in the degree of condensation in the drift models. In a few cases the inclusion of drift increases the mass loss rate, possibly by providing more efficient conditions of wind formation. This behavior is found in a few constant-opacity models where the corresponding PC model shows a low terminal velocity and a mass loss rate at the lower end of the range obtained here.

The effects of drift are, however, more pronounced in the Planck mean models. The winds formed in these models show much lower gas densities, and -- as a consequence of the weaker coupling between the gas and the dust -- show stronger effects of drift. In correspondence with the results in {\rSaHo} we have found several PC models where no drift model wind forms using the same model parameters. All Planck mean PC model winds without a corresponding drift model wind share a low variability, in particular in the degree of condensation. In addition Planck mean drift models tend to be more sensitive to the period of the stellar pulsations than constant-opacity drift (and PC) models are. Concluding we note that drift adds complexity to the wind structure, both in Planck mean models and in constant-opacity models. It is in view of the current results questionable if a simple recipe exists that can reproduce the effects of drift using only PC models.

We have studied how the models in this article compare with the mass loss rate fit formula derived by \citet{WaScWi.:02}. The comparison shows that the mass loss rates are systematically lower than this fit predicts, partly due to the use of different parameters in the models, partly due to the fact that the fit was derived for constant-opacity non-drift models. Planck mean PC models give mass loss rates which are lower than the values of the corresponding constant-opacity models (see above). The fit values are about ten times larger than the computed rates of Planck mean models, with or without drift. The fit formula is in its current form inapplicable to the models presented here.

Although the models in this article are more advanced than previous models they are not yet directly comparable with observations. One essential component, in this context, that is not yet included is frequency dependent radiative transfer \citep[see e.g.\@][]{Ho:99b,HoGaArJo:03}. We expect that the conditions in the wind formation zone of more realistic (non-gray) models will change to somewhere in between the two cases represented by the two different gas opacities studied here, probably closer to the Planck mean models with their lower densities.

Also worth stressing is that we so far only study the short-term evolution in our models (i.e.\@ covering up to a few hundred pulsation periods). A study of long-term evolution, important to stellar evolution for instance, requires longer calculated time-intervals (covering thousands of pulsation periods). We intend to study these aspects closer in future papers. The last point concerns the direct microscopic influence of drift on the dust formation process (nucleation and growth rates). We have so far assumed that the stellar winds will not be significantly affected by this effect (cf.\@ Sect.~\ref{sec:physgdin}). However, we will -- for consistency -- in the near future include these modifications in our models and study the dust formation, affected by drift, in more detail.

\begin{acknowledgements}
All calculations have been performed on the 12-processor HPV9000 placed at the Dept.\@ of Astronomy and Space Physics, financed through a donation by the \emph{Knut and Alice Wallenberg Foundation}. The analysis has been carried out on a SUN workstation financed by the \emph{Swedish Research Council} (VR). This work has been conducted within the framework of the research school on \emph{Advanced Instrumentation and Measurements} (AIM), at Uppsala University. AIM is financially supported by the \emph{Foundation for Strategic Research} (SSF).
\end{acknowledgements}

\bibliographystyle{aa}
\bibliography{/u73/users/christer/Projects/My_Papers/CS_Refs}
\end{document}